\documentclass[preprint,showpacs,preprintnumbers,amsmath,amssymb,showkeys]{revtex4}

\usepackage{graphicx}
\usepackage{dcolumn}
\usepackage{bm}

\usepackage{bm}
\usepackage{braket}
\usepackage{enumitem}
\usepackage[utf8]{inputenc}
\usepackage{amsmath}
\usepackage{amssymb}
\usepackage{amsthm}

\begin{document}


\title{Quantum coherence from Kirkwood-Dirac nonclassicality, some bounds, and operational interpretation}

\author{Agung Budiyono$^{1,2,3}$}
\email{agungbymlati@gmail.com}
\author{Joel F. Sumbowo$^{1,3}$}
\author{Mohammad K. Agusta$^{2,3}$}
\author{Bagus E. B. Nurhandoko$^{4}$}
\affiliation{$^1$Research Center for Quantum Physics, National Research and Innovation Agency, South Tangerang 15314, Republic of Indonesia} 
\affiliation{$^2$Research Center for Nanoscience and Nanotechnology, Bandung Institute of Technology, Bandung, 40132, Republic of Indonesia}
\affiliation{$^3$Department of Engineering Physics, Bandung Institute of Technology, Bandung, 40132, Republic of Indonesia} 
\affiliation{$^4$Department of Physics, Bandung Institute of Technology, Bandung, 40132, Republic of Indonesia} 

\date{\today}

\begin{abstract}
Since the early days, there has been a research program using the nonclassical values of some quasiprobability distributions to indicate the nonclassical aspects of quantum phenomena. In particular, in KD (Kirkwood-Dirac) quasiprobability distribution, the distinctive quantum mechanical feature of noncommutativity which underlies many nonclassical phenomena, manifests in the nonreal values and/or the negative values of the real part. Here, we develop a faithful quantifier of quantum coherence based on the KD nonclassicality which captures simultaneously the nonreality and the negativity of the KD quasiprobability. The KD-nonclassicality coherence thus defined, is upper bounded by the uncertainty of the outcomes of measurement described by a rank-1 PVM (projection-valued measure) corresponding to the incoherent orthonormal basis, quantified by the Tsallis $\frac{1}{2}$-entropy. Moreover, they are identical for pure states so that the KD-nonclassicallity coherence for pure state admits a simple closed expression in terms of measurement probabilities. We then use the Maassen-Uffink uncertainty relation for min-entropy and max-entropy to obtain a lower bound for the KD-nonclassicality coherence of a pure state in terms of optimal guessing probability in measurement described by a PVM noncommuting with the incoherent orthonormal basis. We also derive a trade-off relation for the KD-noncassicality coherences of a pure state relative to a pair of noncommuting orthonormal bases with a state-independent lower bound. Finally, we sketch a variational scheme for a direct estimation of the KD-nonclassicality coherence based on weak value measurement and thereby discuss its relation with quantum contextuality. 
\end{abstract}

\pacs{Valid PACS appear here}
\keywords{quantum coherence, Kirkwood-Dirac nonclassicality, measurement uncertainty, Tsallis $\frac{1}{2}$-entropy, min-entropy, max-entropy, optimal guessing probability, trade-off relation, state-independent lower bound, weak value, contextuality}
\maketitle

\section{Introduction\label{Introduction}} 

Quantum superposition lies at the heart of the nonclassical features of quantum phenomena. It is the source of much conceptual debates about the foundation and meaning of quantum mechanics. The past few decades have witnessed a wealth of works showing that quantum superposition is also an indispensable nonclassical ingredient for the development of a variety of promisingly disruptive schemes of quantum information processing and technology. The simplest manifestation of quantum superposition in a single system is quantum coherence which captures the strength of the superposition in a quantum state relative to some a priori chosen reference orthonormal basis  \cite{Streltsov coherence review}. Equivalently, quantum coherence can also be seen as capturing the failure of commutativity between the quantum state and the reference orthonormal basis \cite{Marvian - Spekkens speakable and unspeakable coherence}. Significant efforts have been made recently to better understand coherence viewed as a resource \cite{Marvian - Spekkens speakable and unspeakable coherence,Aberg quantifying of superposition,Levi quantum coherence measure,Baumgratz quantum coherence measure,Girolami quantum coherence measure,Winter operational resource theory of coherence,Streltsov coherence review,Marvian coherence measure,Chitambar physically consistent resource theory of coherence,Napoli robustness of coherence,Yu alternative resource theory of coherence,Yuan quantum coherence intrinsic randomness} within the general framework of quantum resource theory \cite{Horodecki resource theory,Chitambar resource theories review}. This approach draws inspiration from the remarkable successes of the resource theory of entanglement. Different important measures of coherence have been proposed within the resource theoretical framework many of which having analogous forms to the corresponding entanglement measures. 

On the other hand, the nonclassical values of some quasiprobability distributions have long been traditionally used to signal quantumness or nonclassicality stemming from quantum noncommutativity \cite{Lee quasiprobability review}. Various quasiprobability distributions have been conceived as quantum analogs of phase space distribution in classical statistical mechanics. Quantum noncommutativity implies that no quasiprobability which is convex-linear in the density operator, satisfy all the Kolmogorov axioms for classical probability, i.e., real and nonnegative yielding correct marginal probabilities \cite{Lostaglio KD quasiprobability and quantum fluctuation}. A well-known example is the Wigner function which is real but may assume negative value \cite{Wigner distribution}. In the present work, we consider a specific quasiprobability distribution called KD (Kirkwood-Dirac) quasiprobability \cite{Kirkwood quasiprobability,Dirac quasiprobability,Barut KD quasiprobability,Chaturvedi KD distribution} to quantitatively characterize quantum coherence and study its relation with other nonclassical features of quantum mechanics. KD quasiprobability may take complex value and/or its real part (known as Terletsky-Margenou-Hill quasiprobability \cite{Terletsky TBMH quasiprobability,Barut KD quasiprobability,Margenau TBMH quasiprobability}) may assume negative value, manifesting the noncommutativity between the quantum state and the pair of the defining orthonormal bases. This naturally raises a question: how is the nonreality and/or the negativity of the KD quasiprobability related to the coherence of the associated quantum state relative to a reference orthonormal basis which also captures the noncommutativity between the state and the reference basis? As a partial answer to this question, previously, we have shown that one can indeed quantitatively characterize the coherence of a quantum state on a finite-dimensional Hilbert space by using the nonreality of the associated KD quasiprobability \cite{Agung KD-nonreality coherence}. 

However, quantum noncommutativity does not only manifest in the nonreal values of the KD quasiprobability, but also independently in the negative values of the real part. Moreover, it has been argued over the past decade that the negative values of the real part of the KD quasiprobability, with or without the nonreal part, play crucial roles in different areas of quantum science, such as quantum fluctuation in condensed matter systems \cite{Lostaglio KD quasiprobability and quantum fluctuation,Mehboudi FDT using symmetric logarithmic derivative}, quantum thermodynamics \cite{Allahverdyan TBMH as quasiprobability distribution of work,Lostaglio TBMH quasiprobability fluctuation theorem contextuality,Levy quasiprobability distribution for heat fluctuation in quantum regime}, quantum information scrambling \cite{Alonso KD quasiprobability witnesses quantum scrambling,Halpern quasiprobability and information scrambling}, quantum metrology \cite{Arvidsson-Shukur quantum advantage in postselected metrology,Lupu-Gladstein negativity enhanced quantum phase estimation 2022}, and quantum foundation \cite{Pusey negative TBMH quasiprobability and contextuality,Kunjwal contextuality of non-real weak value,Lostaglio contextuality in quantum linear response}. It is thus desirable to use the negative values of the real part of the KD quasiprobability together with the nonreal part to quantitatively characterize quantum coherence. Here, we argue that the KD nonclassicality, which simultaneously captures the nonreality and the negativity of the KD quasiprobability, can indeed be used to quantitatively characterize quantum coherence. Namely, we construct a faithful quantifier of coherence of a quantum state relative to an incoherent orthonormal basis, based on the KD nonclassicality of the associated KD quasiprobability defined relative to the incoherent basis and another orthonormal basis, and maximized over all possible choices of the latter. 

The proposed coherence quantifier, called KD-nonclassicality coherence, gives a lower bound to the $l_1$-norm coherence \cite{Baumgratz quantum coherence measure}. We then show that the KD-nonclassicality coherence also provides a lower bound to the uncertainty of outcomes of measurement described by a rank-1 PVM (projection-valued measure) corresponding to the incoherent orthonormal basis, quantified by half of the Tsallis entropy with entropy-index $1/2$ \cite{Tsallis on Tsallis entropy}. Moreover, they are shown to be identical for all pure states so that the KD-nonclassicality coherence for pure state admits a simple closed expression in terms of measurement probabilities. It is further upper bounded by the purity of the quantum state $\varrho$ as $\sqrt{d{\rm Tr}(\varrho^2)}-1$, where $d$ is the dimension of the Hilbert space, and for pure states, the maximum, i.e., $\sqrt{d}-1$, is attained by the maximally coherent states. We then use the Maassen-Uffink uncertainty relation \cite{Massen-Uffink entropic UR} for min-entropy and max-entropy to show that the KD-nonclassicality coherence of a pure state can be used to upper bound the optimal guessing probability in the measurement described a rank-1 PVM basis complementary to the incoherent orthonormal basis. Furthermore, we derive a trade-off relation for the KD-nonclassicality coherences of a pure state relative to a pair of orthonormal bases with a state-independent lower bound. We then sketch a variational scheme for a direct estimation of KD nonclassicality coherence using weak value measurement and classical optimization. Thereby, we discuss its relation with proof of quantum contextuality  \cite{Spekkens quantum contextuality} based on weak value measurement using weak measurement with postselection \cite{Pusey negative TBMH quasiprobability and contextuality,Kunjwal contextuality of non-real weak value,Lostaglio contextuality in quantum linear response}. Numerical examples for single and two-qubit states are given.   

The article is organized as follows. In Section \ref{Preliminaries} of Preliminaries, we summarize the concept of quantum coherence relative to a reference orthonormal basis and the KD nonclassicality in KD quasiprobability, highlighting that both nonclassical concepts are different manifestations of quantum noncommutativity. In Section \ref{Quantum coherence via the KD nonclassicality} we define KD-nonclassicality coherence as a faithful coherence quantifier showing that it satisfies certain desirable requirements. In Section \ref{Upper bounds, lower bound and uncertainty relation for KD-nonclassicality coherence} we derive upper bounds for the KD-nonclassicality coherence in terms of $l_1$-norm coherence, measurement uncertainty quantified by Tsallis $\frac{1}{2}$-entropy and state purity. We also derive a lower bound for pure state KD-nonclassicality coherence in terms of optimal guessing probability and noncommutativity, and obtain uncertainty relation for pure state KD-nonclassicality coherences relative to a pair of orthonormal bases with a state-independent lower bound. Finally, in Section \ref{Experimental estimation, proof of quantum contextuality, and static susceptibility}, we sketch its estimation via weak value measurement and discuss its relation with quantum contextuality. We summarize in Section \ref{Summary and Remarks} with concluding remarks. 

\section{Preliminaries\label{Preliminaries}} 

\subsection{Quantum coherence relative to an orthonormal basis}

Quantum coherence captures the degree of superposition of the elements of a reference basis in a quantum state of a single system. Consider a quantum system with a Hilbert space $\mathcal{H}$ of finite dimension $d$, and choose a reference orthonormal basis $\{\ket{a}\}$ of $\mathcal{H}$. We note that the set of projectors $\{\Pi_a:=\ket{a}\bra{a}\}$, $\sum_a\Pi_a=\mathbb{I}$, where $\mathbb{I}$ is the identity operator on $\mathcal{H}$, comprises a rank-1 orthogonal PVM describing a projective measurement. A quantum state represented by a density operator $\varrho$ on $\mathcal{H}$ is incoherent with respect to the reference orthonormal basis $\{\ket{a}\}$ of $\mathcal{H}$ if it can be decomposed as a statistical mixture of the elements of the reference basis, i.e., $\varrho=\sum_a p_a\ket{a}\bra{a}$, where $\{p_a\}$, with $p_a\ge 0$ and $\sum_ap_a=1$, are the mixing probabilities. Hence, the density operator is diagonal in the basis $\{\ket{a}\}$ so that they are commuting, i.e., $[\Pi_a,\varrho]=0$, for all $a$. Any other states, i.e., states with $\varrho\neq\sum_a p_a\ket{a}\bra{a}$, or $[\Pi_a,\varrho]\neq 0$ for some $a$, are coherent relative to the orthonormal basis $\{\ket{a}\}$. In this sense, $\{\ket{a}\}$ is referred to as an incoherent basis.  

In the last decade, motivated by the fact that quantum coherence is a necessary resource in diverse quantum information processing tasks and schemes of quantum technologies, many researchers have applied the framework of quantum resource theory \cite{Horodecki resource theory,Chitambar resource theories review} to rigorously characterize, quantify, and manipulate the coherence in a quantum state \cite{Marvian - Spekkens speakable and unspeakable coherence,Aberg quantifying of superposition,Levi quantum coherence measure,Baumgratz quantum coherence measure,Girolami quantum coherence measure,Winter operational resource theory of coherence,Streltsov coherence review,Marvian coherence measure,Chitambar physically consistent resource theory of coherence,Napoli robustness of coherence,Yu alternative resource theory of coherence,Yuan quantum coherence intrinsic randomness}. In this framework, we are restricted to certain class of incoherence-preserving quantum operations, which are regarded as free. Accordingly, quantum states are divided into two sets. One is the set of free incoherent states, which is mapped by the free operations onto itself. The other is the set of states with coherence that cannot be obtained by applying the free operations to the free incoherent states. Any quantifier of the coherence is then required to be nonincreasing under the free incoherence-preserving operations. There are however several different classes of free operations suggested in the literature, and their operational interpretation are not always clear \cite{Streltsov coherence review,Marvian - Spekkens speakable and unspeakable coherence}. In this work, we will take a more direct approach to quantitatively characterize coherence by connecting it with the notion of quantumness or nonclassicality in KD quasiprobability distribution which has a well-defined interpretation in terms of laboratory operations. Nevertheless, we will show that the quantifier of coherence we propose satisfy certain plausible requirements within a resource theoretical framework. 

\subsection{Kirkwood-Dirac quasiprobability and nonclassicality\label{Kirkwood-Dirac quasiprobability and nonclassicality}}

KD quasiprobability is a specific quantum extension of classical phase space probability distribution applicable for quantum systems with Hilbert space of finite or infinite dimension \cite{Kirkwood quasiprobability,Dirac quasiprobability}. Below, we only consider systems with finite-dimensional Hilbert space. The KD quasiprobability associated with a quantum state $\varrho$ on a finite-dimensional Hilbert space $\mathcal{H}$ over a pair of orthonormal bases $\{\ket{a}\}$ and $\{\ket{b}\}$ of $\mathcal{H}$, is defined as \cite{Kirkwood quasiprobability,Dirac quasiprobability,Barut KD quasiprobability,Chaturvedi KD distribution}
\begin{eqnarray}
{\rm Pr}_{\rm KD}(a,b|\varrho):={\rm Tr}(\Pi_b\Pi_a\varrho). 
\label{KD quasiprobability}
\end{eqnarray}
We refer to $\{\ket{a}\}$ and $\{\ket{b}\}$ respectively as the first and the second defining orthonormal bases. 
KD quasiprobability gives an informationally equivalent representation of an arbitrary quantum state. Namely, given a KD quasiprobability ${\rm Pr}_{\rm KD}(a,b|\varrho)$ defined over a pair of orthonormal bases $\{\ket{a}\}$ and $\{\ket{b}\}$, with $\braket{a|b}\neq 0$ for all $(a,b)$, the associated quantum state can be reconstructed as $\varrho=\sum_{a,b}\braket{a|\varrho|b}\ket{a}\bra{b}=\sum_{a,b}{\rm Pr}_{\rm KD}(a,b|\varrho)\frac{\ket{a}\bra{b}}{\braket{b|a}}$. Hence, ${\rm Pr}_{\rm KD}(a,b|\varrho)/\braket{b|a}$ is the coefficient of the expansion of the density operator with respect to the orthonormal basis $\{\ket{a}\bra{b}\}$ of the space of operators on the Hilbert space $\mathcal{H}$ equipped with Hilbert-Schmidt inner product. 

KD quasiprobability gives correct marginal probabilities, i.e., $\sum_i{\rm Pr}_{\rm KD}(a,b|\varrho)={\rm Tr}({\Pi_j\varrho})={\rm Pr}(j|\varrho)$, $i\neq j$, $i,j=a,b$, where ${\rm Pr}(\cdot)$ is the usual real and nonnegative classical probability. However, and crucially, unlike the conventional classical probability, KD quasiprobability may assume complex value and/or its real part may be negative. Such nonreality and negativity manifest the noncommutativity among $\Pi_a$, $\Pi_b$ and $\varrho$, i.e., assuming two of them commute, renders the KD quasiprobability real and nonnegative. In this sense, the negativity and the nonreality of KD quasiprobability may be seen as a form of quantumness or nonclassicality associated with noncommutativity. It is intriguing that the converse is not necessarily true. Namely, there are cases where pairwise noncommutativity among $\Pi_a$, $\Pi_b$ and $\varrho$ do not imply nonreality and/or negativity of the KD quasiprobability \cite{Drori nonclassicality tighter and noncommutativity,deBievre nonclassicality in KD distribution,deBievre incompatibility-uncertainty-KD nonclassicality,Xu KD classical pure states}. Remarkably, the real and imaginary parts of the KD quasiprobability can be estimated in experiment directly without recoursing to quantum state tomography \cite{Lundeen measurement of KD distribution,Salvail direct measurement KD distribution,Bamber measurement of KD distribution,Thekkadath measurement of density matrix,Johansen quantum state from successive projective measurement,Lostaglio KD quasiprobability and quantum fluctuation,Hernandez-Gomez experimental observation of TBMH negativity,Wagner measuring weak values and KD quasiprobability,Chiribella estimation of weak value,Haapasalo generalized weak value,Vallone strong measurement to reconstruct quantum wave function,Cohen estimating of weak value with strong measurements,Lundeen complex weak value,Jozsa complex weak value}. Recently, this fact has led to a flurry of works using the KD quasiprobability and its nonclassicality in different areas of quantum science. 

KD quasiprobability is always normalized to unity, i.e., $\sum_{a,b}{\rm Pr}_{\rm KD}(a,b|\varrho)=1$, so that we have $\sum_{a,b}|{\rm Pr}_{\rm KD}(a,b|\varrho)|\ge |\sum_{a,b}{\rm Pr}_{\rm KD}(a,b|\varrho)|=1$. It is therefore natural to define the following quantity capturing the nonclassicality in the KD quasiprobabily.
\\
{\bf Definition 1 (KD nonclassicality)}. Given a KD quasiprobability ${\rm Pr}_{\rm KD}(a,b|\varrho)$ associated with a quantum state $\varrho$ on a finite-dimensional Hilbert space $\mathcal{H}$ and a pair of orthonormal bases $\{\ket{a}\}$ and $\{\ket{b}\}$ of $\mathcal{H}$, let us define the KD nonclassicality in the KD quasiprobability as follows \cite{Drori nonclassicality tighter and noncommutativity,Alonso KD quasiprobability witnesses quantum scrambling,Lostaglio KD quasiprobability and quantum fluctuation}:
\begin{eqnarray}
{\rm NCl}(\{{\rm Pr}_{\rm KD}(a,b|\varrho)\})&:=&\sum_{a,b}|{\rm Pr}_{\rm KD}(a,b|\varrho)|-1\nonumber\\
&=&\sum_{a,b}|\braket{b|\Pi_a\varrho|b}|-1.  
\label{KD nonclassicality}
\end{eqnarray}
One can see that the KD nonclassicality ${\rm NCl}(\{{\rm Pr}_{\rm KD}(a,b|\varrho)\})$ is nonnegative by definition, and it vanishes only when the KD quasiprobability ${\rm Pr}_{\rm KD}(a,b|\varrho)$ is real and nonnegative for all $a$ and $b$, so that $|{\rm Pr}_{\rm KD}(a,b|\varrho)|={\rm Pr}_{\rm KD}(a,b|\varrho)$. It thus quantifies the failure of the KD quasiprobability to be both real and nonnegative.   

\section{Quantifying quantum coherence using KD nonclassicality\label{Quantum coherence via the KD nonclassicality}}

Since the KD quasiprobability is an informationally equivalent representation of the quantum state, it is instructive to study the way in which the KD quasiprobability encodes the nonclassical aspects of a quantum state such as coherence, general quantum correlation and entanglement. To this end, the fact that the negativity or/and the nonreality of KD quasiprobability and quantum coherence both capture the noncommutativity between the quantum state and the set of projectors associated with certain orthonormal bases of the Hilbert space, naturally gives rise to a question if the former can be used to quantitatively characterize the latter. To answer this question, we must note that the KD quasiprobability is defined based on a pair of orthonormal bases while coherence is defined relative to a single reference orthonormal basis.  

In the previous work \cite{Agung KD-nonreality coherence}, we have argued that the total sum of the nonreality of the KD quasiprobability maximized over one of the two defining orthonormal bases, can be used to quantify the coherence in the state relative to the other defining basis (see Eq. (\ref{KD-nonreality coherence})). On the other hand, the negative values of the real part of the KD quasiprobability also capture the quantum noncommutativity independently of the nonreal values. Furthermore, as mentioned in the Introduction, the negativity of the KD quasiprobability has been used to indicate quantumness in different areas of quantum science and technology independently of the nonreal part \cite{Lostaglio KD quasiprobability and quantum fluctuation,Mehboudi FDT using symmetric logarithmic derivative,Allahverdyan TBMH as quasiprobability distribution of work,Lostaglio TBMH quasiprobability fluctuation theorem contextuality,Levy quasiprobability distribution for heat fluctuation in quantum regime,Alonso KD quasiprobability witnesses quantum scrambling,Halpern quasiprobability and information scrambling,Arvidsson-Shukur quantum advantage in postselected metrology,Lupu-Gladstein negativity enhanced quantum phase estimation 2022,Pusey negative TBMH quasiprobability and contextuality,Kunjwal contextuality of non-real weak value,Lostaglio contextuality in quantum linear response}. It is therefore instructive to ask if the KD nonclassicality defined in Eq. (\ref{KD nonclassicality}), which takes into account simultaneously the nonreality and the negativity of the KD quasiprobability, provides a more complete characterization of the coherence of the associated quantum state. We argue in this section that one can indeed devise a faithful quantifier of coherence in terms of the KD nonclassicality. 

{\bf Definition 2 (KD-nonclassicality coherence)}. Given a general quantum state $\varrho$ on a finite-dimensional Hilbert space $\mathcal{H}$ and an incoherent reference orthonormal basis $\{\ket{a}\}$ of $\mathcal{H}$, we define the following quantity which maps the quantum state and the incoherent orthonormal basis to a nonnegative real number:
\begin{eqnarray}
C_{\rm KD}^{\rm NCl}(\varrho;\{\Pi_a\})&:=&\sup_{\{\ket{b}\}\in\mathcal{B}_{\rm o}(\mathcal{H})}{\rm NCl}(\{{\rm Pr}_{\rm KD}(a,b|\varrho)\})\nonumber\\
&=&\sup_{\{\ket{b}\}\in\mathcal{B}_{\rm o}(\mathcal{H})}\sum_{a,b}\big|\braket{b|\Pi_a\varrho|b}\big|-1, 
\label{KD-nonclassicality coherence}
\end{eqnarray}
where the supremum is taken over the set $\mathcal{B}_{\rm o}(\mathcal{H})$ of all the othornormal bases of $\mathcal{H}$. 

{\bf Remark}. Consider a composite of $N$ subsystems and a product incoherent orthonormal basis $\{\ket{a}\}=\{\ket{a_1}\otimes\cdots\otimes\ket{a_N}\}:=\{\ket{a_1,\dots,a_N}\}$, where $\{\ket{a_i}\}$ is an orthonormal basis for the Hilbert space $\mathcal{H}_i$ of subsystem $i$,  relative to which we wish to quantify the coherence of the $N$-partite quantum state $\varrho_{1\cdots N}$ on the Hilbert space $\mathcal{H}=\otimes_{i=1}^N\mathcal{H}_i$. Hence, $\{\ket{a}\}$ is the first orthonormal basis for defining the KD quasiprobability ${\rm Pr}_{\rm KD}(a,b|\varrho)$. In this case, we then assume that the second basis for defining the KD quasiprobability in Eq. (\ref{KD-nonclassicality coherence}) is also a product, i.e., $\{\ket{b}\}=\{\ket{b_1,\dots,b_N}\}$ where $\{\ket{b_i}\}$ is the second orthonormal basis of $\mathcal{H}_i$ of subsystem $i$. 

We argue that $C_{\rm KD}^{\rm NCl}(\varrho;\{\Pi_a\})$ defined in Eq. (\ref{KD-nonclassicality coherence}) possesses the following desirable properties for a quantifier of quantum coherence of $\varrho$ relative to the incoherent orthonormal  basis $\{\ket{a}\}$. 

\begin{enumerate}[label=(\roman*)]

\item {\it Faithfulness}, i.e., $C_{\rm KD}^{\rm NCl}(\varrho;\{\Pi_a\})=0$ if and only if the quantum state $\varrho$ is incoherent with respect to the orthonormal basis $\{\ket{a}\}$. 

\item {\it Unitary covariance}, i.e., $C_{\rm KD}^{\rm NCl}(U\varrho U^{\dagger};\{ U\Pi_a U^{\dagger}\})=C_{\rm KD}^{\rm NCl}(\varrho;\{\Pi_a\})$ for arbitrary unitary transformation $U$.

\item {\it Convexity}, i.e., $C_{\rm KD}^{\rm NCl}(\sum_kp_k\varrho_k;\{\Pi_a\})\le\sum_kp_kC_{\rm KD}^{\rm NCl}(\varrho_k;\{\Pi_a\})$, where $\{p_k\}$ is a set of mixing probabilities: $0\le p_k\le 1$, $\sum_kp_k=1$.

\item {\it Nonincreasing under partial trace}, i.e., $C_{\rm KD}^{\rm NCl}(\varrho_{12};\{\Pi_{a_1}\otimes\mathbb{I}_2\})\ge C_{\rm KD}^{\rm NCl}(\varrho_1;\{\Pi_{a_1}\})$, where $\varrho_{12}$ is a bipartite state on the Hilbert space $\mathcal{H}=\mathcal{H}_1\otimes\mathcal{H}_2$, $\varrho_1={\rm Tr}_2\varrho_{12}$, $\{\ket{a_1}\}$ is the incoherent orthonormal basis of the Hilbert space $\mathcal{H}_1$ of system 1, and $\mathbb{I}_2$ is the identity operator on the Hilbert space $\mathcal{H}_2$ of system 2. Moreover, equality is attained when the bipartite state takes a product form: $\varrho_{12}=\varrho_1\otimes\varrho_2$. Property (iv) captures the intuition that if two systems are correlated, ignoring one of them should not increase the coherence of the other. Note that partial trace is regarded as a free operation in quantum resource theory \cite{Chitambar resource theories review}.

\item {\it Nonincreasing under decoherence operation}, i.e., $C_{\rm KD}^{\rm NCl}(\varrho;\{\Pi_a\})\ge C_{\rm KD}^{\rm NCl}(\varrho';\{\Pi_a\})$, where $\varrho'=p\varrho+(1-p)\mathcal{D}(\varrho;\{\Pi_a\})$, $0\le p\le 1$, and $\mathcal{D}(\varrho;\{\Pi_a\}):=\sum_a\Pi_a\varrho\Pi_a$ is the dephasing operation (measurement channel) which removes the off-diagonal terms of the density operator $\varrho$ in the basis $\{\ket{a}\}$. 

\item {\it Nonincreasing under coarsegraining}. Consider a coarsegraining of the incoherent basis and define a coarse-grained KD quasiprobability as follows ${\rm Pr}_{\rm KD}(A,b|\varrho):=\sum_{a\in A}{\rm Pr}_{\rm KD}(a,b|\varrho)={\rm Tr}(\Pi_b\Pi_A\varrho)$, where $\{A\}$ is a disjoint subset partitioning of the set $\{a\}$ and $\Pi_A:=\sum_{a\in A}\Pi_a$. Then,  one has $C_{\rm KD}^{\rm NCl}(\varrho;\{\Pi_A\}):=\sup_{\{\ket{b}\}\in\mathcal{B}_{\rm o}(\mathcal{H})}\sum_{A,b}|{\rm Pr}_{\rm KD}(A,b|\varrho)|-1\le C_{\rm KD}^{\rm NCl}(\varrho;\{\Pi_a\})$. 

\item {\it Nonincreasing under a controlled incoherent permutation of the element of the incoherent basis}, i.e., $C_{\rm KD}^{\rm NCl}(\Phi_{\rm CIP}(\varrho_S);\{\Pi_a\})\le C_{\rm KD}^{\rm NCl}(\varrho_S;\{\Pi_a\})$. Here, $\Phi_{\rm CIP}(\varrho_S)={\rm Tr}_E\big(U_{SE}(\varrho_S\otimes\varrho_E)U_{SE}^{\dagger}\big)$, where $\varrho_E$ is the state of an ancilla (auxiliary system) which is assumed to be incoherent relative to some a priori fixed basis $\{\ket{e}_E\}$, $U_{SE}$ is a unitary operator defined as: $U_{SE}\ket{a}_S\ket{e}_E=\ket{\mu_e(a)}_S\ket{e}_E$, where $\mu_e(a)$ is a permutation of the elements of the incoherent basis $\{\ket{a}\}$ of the principal system conditioned on the element of the incoherent basis of the ancilla, and the partial trace is taken over the ancilla. We show in Appendix \ref{Proofs of Properties (i)-(vi)} that such a quantum channel is incoherence preserving.  
\end{enumerate}

We sketch the proofs of properties (i)-(vii) in Appendix \ref{Proofs of Properties (i)-(vi)}. We note that we have not been able to prove the monotonicity of $C_{\rm KD}^{\rm NCl}(\varrho;\{\Pi_a\})$ with respect to the different classes of incoherent operations suggested in the literature \cite{Streltsov coherence review}. Rather, as stated in the property (vii), we showed that it is monotonic under a more restricted class of operations of controlled incoherent permutation. Here on, we shall call $C_{\rm KD}^{\rm NCl}(\varrho;\{\Pi_a\})$ defined in Eq. (\ref{KD-nonclassicality coherence}) the KD-nonclassicality coherence of the state $\varrho$ relative to the incoherent orthonormal basis $\{\ket{a}\}$. 

We note that the computation of the KD-nonclassicality coherence of a generic state involves optimization which in general is analytically intractable. However, for all pure states, as will be shown in the next Section, it admits a simple closed expression in terms of measurement uncertainty given by the Tsallis $\frac{1}{2}$-entropy. This situation is similar to e.g. the relative entropy of coherence \cite{Baumgratz quantum coherence measure} which has a simple expression in terms of measurement Shannon entropy for pure states, but for mixed state we need to make diagonalization of the density matrix which is in general analytically intractable. 

Let us give an illustration of the numerical computation of the KD-nonclassicality coherence of a single qubit. See Appendix \ref{Numerical examples of Proposition 3 for single and two qubits} for numerical examples of two-qubits. Without loss of generality, we take as the incoherent orthonormal basis: $\{\ket{a}\}=\{\ket{0},\ket{1}\}$, i.e., the set of the eigenvectors of the Pauli operator $\sigma_z$. It is convenient to express the general density operator of the qubit as $\varrho=(\mathbb{I}+\vec{r}\cdot\vec{\sigma})/2$, where $\vec{r}=(r_x,r_y,r_z)^{\rm T}$ is a vector satisfying $r_x^2+r_y^2+r_z^2=r^2\le 1$ and $\vec{\sigma}=(\sigma_x,\sigma_y,\sigma_z)^{\rm T}$ is a vector of $x,y,z$ Pauli operators. Using the Bloch sphere parameterization, the density matrix of a qubit with respect to the incoherent basis in general reads as 
\begin{eqnarray}
&&\{\braket{a|\varrho|a'}\}_{aa'}\nonumber\\
&=&
\begin{pmatrix} 
\frac{1+r\cos\theta}{2} & \frac{r(\sin\theta\cos\varphi-i\sin\theta\sin\varphi)}{2}&\\
\frac{r(\sin\theta\cos\varphi+i\sin\theta\sin\varphi)}{2} & \frac{1-r\cos\theta}{2}&
\end{pmatrix},
\label{density matrix for a single qubit in Bloch sphere}
\end{eqnarray}
where $a,a'=\{0,1\}$, $\theta\in[0,\pi]$ and $\varphi\in[0,2\pi)$ are the polar and azimuthal angles respectively, and $r$ characterizes the purity of the state as: $\gamma:={\rm Tr}(\varrho^2)=\frac{1+r^2}{2}$. 

\begin{figure}[h]
\centering
\includegraphics[width=.9\columnwidth]{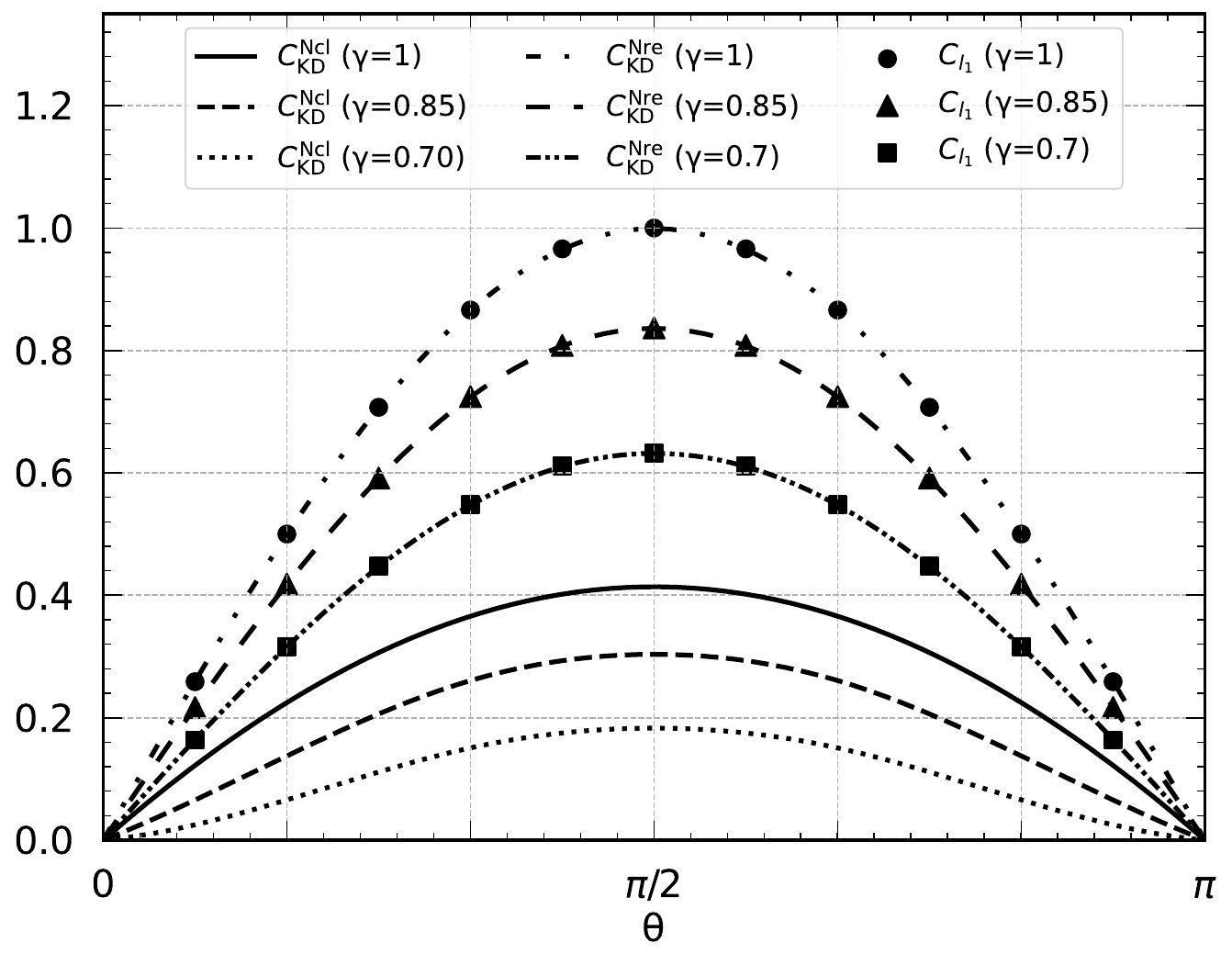}
\caption{\label{fig:epsart}KD-nonclassicality coherence $C_{\rm KD}^{\rm NCl}(\varrho;\{\Pi_a\})$ with $\{\ket{a}\}=\{\ket{0},\ket{1}\}$, of a single qubit with the density matrix of Eq. (\ref{density matrix for a single qubit in Bloch sphere}) and $\varphi=0$, for different values of purity $\gamma={\rm Tr}(\varrho^2)$, as a function of $\theta$. We also plotted the $l_1$-norm coherence $C_{l_1}(\varrho;\{\Pi_a\})$ defined in Eq. (\ref{l1-norm coherence}) and KD-nonreality coherence $C_{\rm KD}^{\rm NRe}(\varrho;\{\Pi_a\})$ defined in Eq. (\ref{KD-nonreality coherence}). For the KD-nonclassicality coherence, the peak saturates the upper bound in Eq. (\ref{KD-nonclassicality coherence is upper bounded by purity step 1}), i.e., $\sqrt{2{\rm Tr}(\varrho^2)}-1$, at $\theta=\pi/2$. See the main text for detail.}
\end{figure} 

Next, for the purpose of computation, we express the second basis $\{\ket{b(\alpha,\beta)}\}=\{\ket{b_+(\alpha,\beta)},\ket{b_-(\alpha,\beta)}\}$ for defining the KD quasiprobability using the Bloch sphere parameterization as: 
\begin{eqnarray}
\ket{b_+(\alpha,\beta)}&:=&\cos\frac{\alpha}{2}\ket{0}+\sin\frac{\alpha}{2}e^{i\beta}\ket{1};\nonumber\\
\ket{b_-(\alpha,\beta)}&:=&\sin\frac{\alpha}{2}\ket{0}-\cos\frac{\alpha}{2}e^{i\beta}\ket{1}, 
\label{complete set of bases for two-dimensional Hilbert space}
\end{eqnarray}
with $\alpha\in[0,\pi]$, $\beta\in[0,2\pi)$. Hence, upon varying the angular parameters $\alpha$ and $\beta$ over the whole ranges of their values, one scans all the possible orthonormal bases $\{\ket{b(\alpha,\beta)}\}\in\mathcal{B}_{\rm o}(\mathbb{C}^2)$ of the two-dimensional Hilbert space $\mathbb{H}\cong\mathbb{C}^2$. The KD-nonclassicality coherence of the qubit relative to the incoherent basis $\{\ket{a}\}=\{\ket{0},\ket{1}\}$ in then obtained by inserting Eqs. (\ref{density matrix for a single qubit in Bloch sphere}) and (\ref{complete set of bases for two-dimensional Hilbert space}) into Eq. (\ref{KD-nonclassicality coherence}) and solving the optimization problem. In Fig. 1, we plotted the KD-nonclassicality coherence of a single qubit with the density matrix $\braket{a|\varrho|a'}$ having the form of Eq. (\ref{density matrix for a single qubit in Bloch sphere}) and $\varphi=0$ as a function of the polar angle $\theta\in[0,\pi]$, for different values of purity, obtained numerically following the above procedure. As expected, KD-nonclassicality coherence is vanishing for incoherent states, i.e., $\theta=0,\pi$, and it takes maximum value for $\theta=\pi/2$ which for pure state, i.e., $\gamma=1$, corresponds to the maximally coherent state. All the graphs in Fig. 1 for the KD-nonclassicality coherence do not change as we vary $\varphi\in[0,2\pi)$

For comparison, in Fig. 1, we plotted the corresponding $l_1$-norm coherence defined  as \cite{Baumgratz quantum coherence measure} 
\begin{eqnarray}
C_{l_1}(\varrho;\{\Pi_a\}):=\sum_{a\neq a'}|\braket{a|\varrho|a'}|. 
\label{l1-norm coherence}
\end{eqnarray}
The $l_1$-norm coherence is mathematically the most intuitive measure of coherence quantifying directly the magnitude of the off-diagonal terms of the density matrix reflecting the strength of the superposition between the elements of the incoherent basis. We also plotted the coherence quantifier based on the maximum nonreality of the KD quasiprobability proposed in Ref. \cite{Agung KD-nonreality coherence}, here referred to as KD-nonreality coherence, defined as 
\begin{eqnarray}
C_{\rm KD}^{\rm NRe}(\varrho;\{\Pi_a\}):=\sup_{\{\ket{b}\}\in\mathcal{B}_{\rm o}(\mathcal{H})}{\rm NRe}(\{{\rm Pr}_{\rm KD}(a,b|\varrho)\}), 
\label{KD-nonreality coherence}
\end{eqnarray}
where ${\rm NRe}(\{{\rm Pr}_{\rm KD}(a,b|\varrho)\}):=\sum_{a,b}|{\rm Im}{\rm Pr}_{\rm KD}(a,b|\varrho)|$ is the KD nonreality in the KD quasiprobability ${\rm Pr}_{\rm KD}(a,b|\varrho)$. We showed in Ref. \cite{Agung KD-nonreality coherence} that for arbitrary state of a single qubit, we have $C_{\rm KD}^{\rm NRe}(\varrho;\{\Pi_a\})=C_{l_1}(\varrho;\{\Pi_a\})$, which for the value of parameters considered, can be seen in Fig. 1. One also finds in Fig. 1 that the $l_1$-norm coherence is never less than the KD-nonclassicality coherence. This in fact applies for general state on any finite-dimensional Hilbert space as shown in the next Section.    

\section{Some bounds \label{Upper bounds, lower bound and uncertainty relation for KD-nonclassicality coherence}}

\subsection{Upper bounds: $l_1$-norm coherence, measurement uncertainty, state purity\label{Upper bounds: l_1-norm coherence, measurement uncertainty, state purity}}

Let us proceed to develop several upper bounds for the KD-nonclassicality coherence.  

First, we have the following proposition.\\
{\bf Proposition 1}. The KD-nonclassicality coherence of a state $\varrho$ on a finite-dimensional Hilbert space $\mathcal{H}$ relative to an incoherent orthonormal basis $\{\ket{a}\}$ of $\mathcal{H}$ gives a lower bound to the corresponding $l_1$-norm coherence defined in Eq. (\ref{l1-norm coherence}), i.e., 
\begin{eqnarray}
C_{\rm KD}^{\rm NCl}(\varrho ;\{\Pi_a\})\le C_{l_1}(\varrho;\{\Pi_a\}). 
\label{KD-nonclassicality coherence is upper bounded by l1-norm coherence}
\end{eqnarray}
{\bf Proof}. The proof is straightforward given in Appendix \ref{Proof of Proposition 1}

Next, recall that the KD-nonclassicality coherence $C_{\rm KD}^{\rm NCl}(\varrho;\{\Pi_a\})$ captures the noncommutativity between the state $\varrho$ and the rank-1 PVM $\{\Pi_a\}$ associated with the incoherent orthonormal basis $\{\ket{a}\}$, i.e., there is a subset of $\{a\}$ such that $[\Pi_a,\varrho]\neq 0$. Such noncommutativity must lead to uncertainty of outcomes of the measurement described by the PVM basis $\{\Pi_a\}$ over the state $\varrho$. Since the measurement uncertainty must also take into account the classical uncertainty originating from the statistical mixing in the preparation of the state $\varrho$ when it is not pure, the measurement uncertainty must somehow give another upper bound to the KD-nonclassicality coherence. The following proposition corroborates the above intuition.\\
{\bf Proposition 2}. The KD-nonclassicality coherence of a quantum state $\varrho$ on a Hilbert space $\mathcal{H}$ with a finite dimension $d$ relative to an incoherent orthonormal basis $\{\ket{a}\}$ of $\mathcal{H}$ is upper bounded by the uncertainty of the outcomes of measurement described by the PVM $\{\Pi_a\}$ over the state $\varrho$ as
\begin{eqnarray}
\label{KD negativity is upper bounded by measurement uncertainty 1}
C_{\rm KD}^{\rm NCl}(\varrho ;\{\Pi_a\})&\le&\sum_a\sqrt{{\rm Pr}(a|\varrho)}-1\\
\label{KD negativity is upper bounded by measurement uncertainty 2}
&=&\frac{1}{2}S_{\frac{1}{2}}(\{{\rm Pr}(a|\varrho)\})\\
\label{KD negativity is upper bounded by measurement uncertainty 3}
&\le&\sqrt{d}-1,
\end{eqnarray}
where $S_{\frac{1}{2}}(\{{\rm Pr}(a|\varrho)\}):=2(\sum_a\sqrt{{\rm Pr}(a|\varrho)}-1)$ is the Tsallis $\frac{1}{2}$-entropy capturing the randomness of drawing samples from the probability ${\rm Pr}(a|\varrho)$ \cite{Tsallis on Tsallis entropy}. Moreover, for pure states, the inequality in Eq. (\ref{KD negativity is upper bounded by measurement uncertainty 1}) becomes an equality. \\
{\bf Proof}. We first have, from the definition of the KD-nonclassicality coherence in Eq. (\ref{KD-nonclassicality coherence}), 
\begin{eqnarray}
&&C_{\rm KD}^{\rm NCl}(\varrho ;\{\Pi_a\})+1\nonumber\\
&=&\sup_{\{\ket{b}\}\in\mathcal{B}_{\rm o}(\mathcal{H})}\sum_a\sum_b\Big|\frac{{\rm Tr}(\Pi_b\Pi_a\varrho)}{{\rm Tr}(\Pi_b\varrho)}\Big|{\rm Tr}(\Pi_b\varrho)\nonumber\\
\label{from weak measurement to quantum uncertainty step 1}
&\le&\sum_a\Big(\sum_{b_*}\Big|\frac{{\rm Tr}(\Pi_{b_*}\Pi_a\varrho)}{{\rm Tr}(\Pi_{b_*}\varrho)}\Big|^2{\rm Tr}(\Pi_{b_*}\varrho)\Big)^{1/2}\\
\label{from weak measurement to quantum uncertainty step 2}
&=&\sum_a\Big(\sum_{b_*}\frac{|{\rm Tr}(\Pi_{b_*}\Pi_a\varrho )|^2}{{\rm Tr}(\Pi_{b_*}\varrho )}\Big)^{1/2},
\end{eqnarray}
where $\{\ket{b_*}\}\in\mathcal{B}_{\rm o}(\mathcal{H})$ in Eq. (\ref{from weak measurement to quantum uncertainty step 1}) is a second basis which achieves the supremum and we have also used the Jensen inequality. Next, applying the Cauchy-Schwartz inequality to the numerator on the right-hand side of Eq. (\ref{from weak measurement to quantum uncertainty step 2}), i.e., $|{\rm Tr}(\Pi_{b_*}\Pi_a\varrho )|^2=|{\rm Tr}((\Pi_{b_*}^{1/2}\Pi_a\varrho ^{1/2})(\varrho ^{1/2}\Pi_{b_*}^{1/2}))|^2\le{\rm Tr}(\Pi_{b_*}\Pi_a\varrho \Pi_a){\rm Tr}(\varrho \Pi_{b_*})$, and using the completeness relation $\sum_{b_*}\Pi_{b_*}=\mathbb{I}$, we obtain
\begin{eqnarray}
C_{\rm KD}^{\rm NCl}(\varrho ;\{\Pi_a\})&\le&\sum_a\big({\rm Tr}(\Pi_a^2\varrho )\big)^{1/2}-1\nonumber\\
\label{KD negativity versus uncertainty step 3}
&=&\sum_a\sqrt{{\rm Pr}(a|\varrho)}-1\nonumber\\
\label{KD negativity versus uncertainty step 4}
&=&\frac{1}{2}S_{\frac{1}{2}}(\{{\rm Pr}(a|\varrho)\}),\nonumber 
\end{eqnarray}   
where we have used ${\rm Tr}(\Pi_a^2\varrho)={\rm Tr}(\Pi_a\varrho)={\rm Pr}(a|\varrho)$. Since the maximum of the entropy is obtained when ${\rm Pr}(a|\varrho)=1/d$, one has, upon inserting into Eq. (\ref{KD negativity is upper bounded by measurement uncertainty 1}), $C_{\rm KD}^{\rm NCl}(\varrho ;\{\Pi_a\})\le\sqrt{d}-1$. 

To prove the second half of the proposition, first, for pure states $\varrho=\ket{\psi}\bra{\psi}$, the KD-nonclassicality coherence reads
\begin{eqnarray}
&&C_{\rm KD}^{\rm NCl}(\ket{\psi}\bra{\psi};\{\Pi_a\})\nonumber\\
&=&\sup_{\{\ket{b}\}\in\mathcal{B}_{\rm o}(\mathcal{H})}\sum_{a,b}\big|\braket{b|a}\braket{a|\psi}\braket{\psi|b}\big|-1. 
\label{KD-nonclassicality coherence for pure states}
\end{eqnarray}
Next, for a Hilbert space with any finite dimension $d$, it is always possible to find a triple of mutually unbiased bases (MUB) \cite{Durt MUB review}. Let then $\ket{\psi}$ be an element of a basis, say $\{\ket{c}\}$. Then, it is always possible to find a basis $\{\ket{b_*}\}$ which is mutually unbiased with both $\{\ket{a}\}$ and $\{\ket{c}\}$. In this case, we have $|\braket{\psi|b_*}|=|\braket{b_*|a}|=1/\sqrt{d}$, so that inserting these into Eq. (\ref{KD-nonclassicality coherence for pure states}) we get
\begin{eqnarray}
C_{\rm KD}^{\rm NCl}(\ket{\psi}\bra{\psi};\{\Pi_a\})&=&\sum_{a}\big|\braket{a|\psi}\big|-1\nonumber\\
&=&\sum_a\sqrt{{\rm Pr}(a|\ket{\psi}\bra{\psi})}-1. 
\label{KD-nonclassicality coherence is equal to the measurement uncertainty for pure states}
\end{eqnarray}   
\qed 

\begin{figure}[h]
\centering
\includegraphics[width=.9\columnwidth]{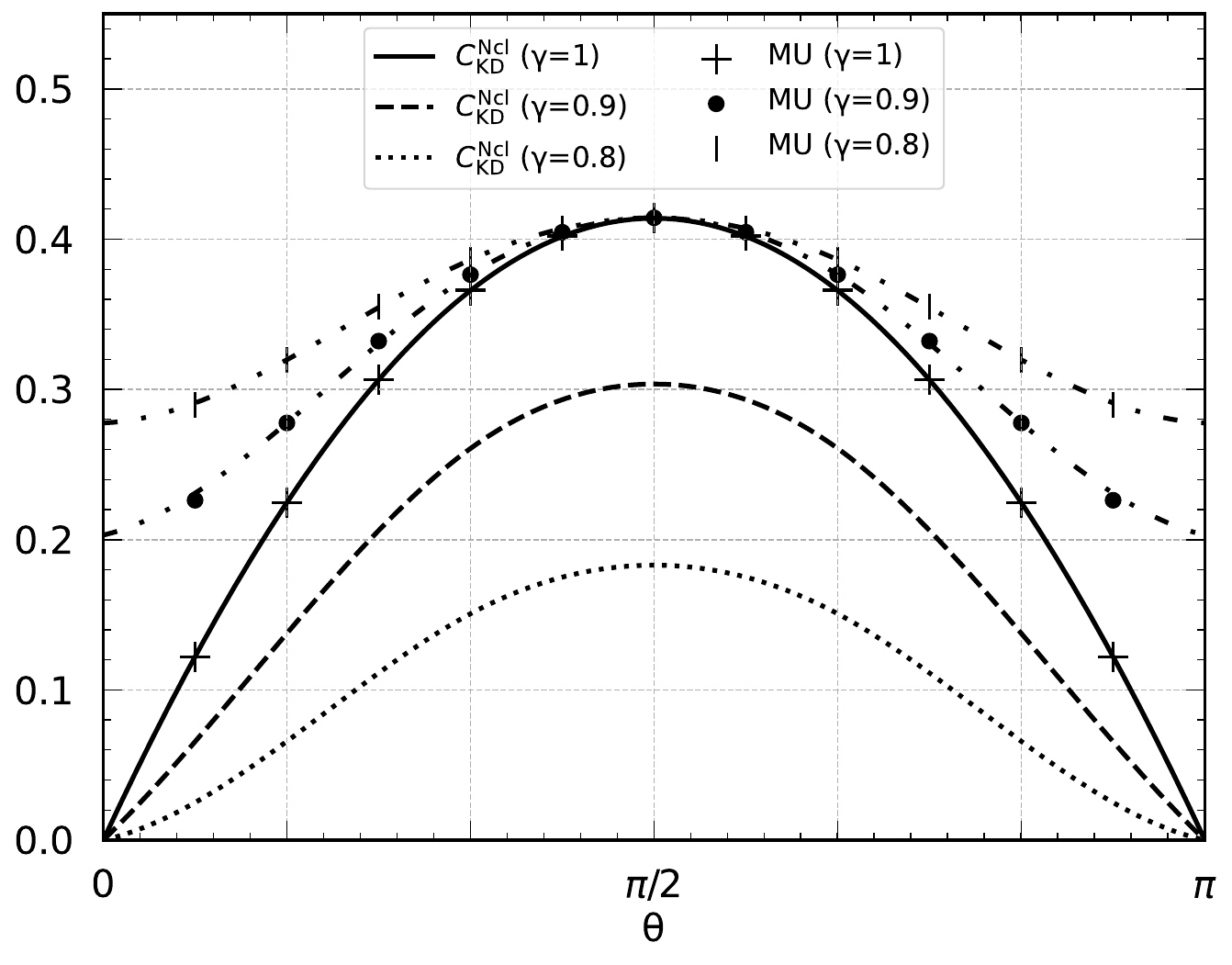}
\caption{\label{fig:epsart}KD-nonclassicality coherence $C_{\rm KD}^{\rm NCl}(\varrho ;\{\Pi_a\})$ with $\{\ket{a}\}=\{\ket{0},\ket{1}\}$, for a single qubit with the density matrix of Eq. (\ref{density matrix for a single qubit in Bloch sphere}) and $\varphi=0$, versus the measurement uncertainty (MU): $\frac{1}{2}S_{\frac{1}{2}}(\{{\rm Pr}(a|\varrho)\})=\sum_a\sqrt{{\rm Pr}(a|\varrho)}-1$, for different values of purity $\gamma$. See the main text for detail.}
\end{figure} 

Hence, the KD-nonclassicality coherence indeed captures measurement uncertainty, in the sense that it gives a lower bound to the latter that is quantified by half of the Tsallis entropy with entropy index $1/2$. Moreover, for pure states $\varrho=\ket{\psi}\bra{\psi}$, they are identical so that in this case we have a simple closed expression of the KD-nonclassicality coherence in terms of probability of measurement given in Eq. (\ref{KD-nonclassicality coherence is equal to the measurement uncertainty for pure states}). In Fig. 2, we plotted the KD-nonclassicality coherence $C_{\rm KD}^{\rm NCl}(\varrho ;\{\Pi_a\})$ of a single qubit relative to the incoherent basis $\{\ket{a}\}=\{\ket{0},\ket{1}\}$, together with the measurement uncertainty $\frac{1}{2}S_{\frac{1}{2}}(\{{\rm Pr}(a|\varrho)\})=\sum_a\sqrt{{\rm Pr}(a|\varrho)}-1$, for the density matrix given by Eq. (\ref{density matrix for a single qubit in Bloch sphere}) with $\varphi=0$ and several values of state purity, as a function of $\theta\in[0,\pi]$, confirming Eq. (\ref{KD negativity is upper bounded by measurement uncertainty 1}). One can see that for pure state with $\gamma=1$, the KD-nonclassicality coherence  is indeed equal to  the measurement uncertainty in accord with Proposition 2.  

Note however that the measurement uncertainty also takes into account the contribution from the classical uncertainty originating from the statistical mixing of the state when it is not pure, so that it is not vanishing for any incoherent mixed state. For a single qubit, this can be seen in Fig. 2 for $\gamma < 1$ when $\theta=0,\pi$. As an extreme example, one can see that for both the maximally coherent state in $d$-dimensional Hilbert space, i.e., $\ket{\psi_{\rm mc}}=\frac{1}{\sqrt{d}}\sum_ae^{i\theta_a}\ket{a}$, $\theta_a\in\mathbb{R}$, and the maximally incoherent state, i.e., the maximally mixed state $\varrho_{\rm mm}=\mathbb{I}/d$, the upper bound in Eq. (\ref{KD negativity is upper bounded by measurement uncertainty 1}) are equal to $\sqrt{d}-1$ saturating the inequality in Eq. (\ref{KD negativity is upper bounded by measurement uncertainty 3}). Hence, measurement uncertainty quantified by the Tsallis $\frac{1}{2}$-entropy cannot distinguish coherent state from incoherent state. By contrast, the KD-nonclassicality coherence vanishes for incoherent states and nonvanishing for coherent states. 

The above observation suggests that it is desirable to have an upper bound in terms of the state purity, which captures the intuition that the coherence should decrease when the purity decreases. Indeed, for the case of a single qubit, this intuition is corroborated in Fig. 2. Namely, as we decrease the purity, the KD-nonclassicality coherence also decreases, but the measurement uncertainty increases. We obtain the following general result.  \\
{\bf Proposition 3}. The KD-nonclassicality coherence of a state $\varrho$ on a Hilbert space $\mathcal{H}$ with a finite dimension $d$ relative to any incoherent orthonormal basis $\{\ket{a}\}$ of $\mathcal{H}$ is bounded from above as 
\begin{eqnarray}
\label{KD-nonclassicality coherence is upper bounded by purity step 1}
C_{\rm KD}^{\rm NCl}(\varrho;\{\Pi_a\})&\le& d^{1/2}{\rm Tr}(\varrho^2)^{1/2}-1\\
\label{KD-nonclassicality coherence is upper bounded by purity step 2}
&\le&\sqrt{d}-1. 
\end{eqnarray}
The upper bound in Eq. (\ref{KD-nonclassicality coherence is upper bounded by purity step 1}) is attained when the second orthonormal basis $\{\ket{b_*}\}$ which achieves the supremum in Eq. (\ref{KD-nonclassicality coherence}) is mutually unbiased with the incoherent orthonormal basis $\{\ket{a}\}$. For a pure state, the upper bound in Eq. (\ref{KD-nonclassicality coherence is upper bounded by purity step 2}) is attained for maximally coherent states: $\ket{\psi_{\rm mc}}=\frac{1}{\sqrt{d}}\sum_ae^{i\theta_a}\ket{a}$, $\theta_a\in\mathbb{R}$. Moreover, $\ket{\psi_{\rm mc}}$ is also maximally coherent relative to $\{\ket{b_*}\}$. \\
{\bf Proof}. From the definition of the KD-nonclassicality coherence, we directly have 
\begin{eqnarray}
&&C_{\rm KD}^{\rm NCl}(\varrho;\{\Pi_a\})\nonumber\\
&=&\sup_{\{\ket{b}\}\in\mathcal{B}_{\rm o}(\mathcal{H})}\sum_{a,b}|\braket{a|\varrho|b}||\braket{b|a}|-1\nonumber\\
\label{the absolute sum of KD nonclassicality and purity step 1}
&\le&d^{1/2}\big(\sum_{a,b_*}|\braket{a|\varrho|b_*}|^2\big)^{1/2}-1\\
\label{the absolute sum of KD nonclassicality and purity step 2}
&=&d^{1/2}{\rm Tr}(\varrho^2)^{1/2}-1,
\end{eqnarray}
where $\{\ket{b_*}\}$ is a basis which achieves the supremum, and we have made use of the Cauchy-Schwartz inequality to get Eq. (\ref{the absolute sum of KD nonclassicality and purity step 1}) and the completeness relation for the orthonormal basis to get Eq. (\ref{the absolute sum of KD nonclassicality and purity step 2}). The inequality in Eq. (\ref{KD-nonclassicality coherence is upper bounded by purity step 2}) is then obtained by noting the fact that the purity of a quantum state $\varrho$ is not larger than 1, i.e., ${\rm Tr}(\varrho^2)\le 1$.  

Further, the condition for equality in Eq. (\ref{the absolute sum of KD nonclassicality and purity step 1}), which is obtained using the Cauchy-Schwartz inequality, is $|\braket{a|\varrho|b_*}|=\eta_{\varrho}|\braket{a|b_*}|$ for some scalar $\eta_{\varrho}$ which is independent of the index $(a,b_*)$. This can only be true if $|\braket{a|\varrho|b_*}|$ is factorizable into a function of $(a,b_*)$ and that of $\varrho$, or, if $|\braket{a|\varrho|b_*}|$ is a uniform function independent of $a$ and $b_*$. Since the former condition does not apply, the latter must be true. But in this case, $|\braket{a|b_*}|$ must also be a uniform function independent of $(a,b_*)$. To guarantee normalization, we thus have $|\braket{a|b_*}|=1/\sqrt{d}$ for all $a,b_*$, which means that the pair of bases $\{\ket{a}\}$ and $\{\ket{b_*}\}$ are mutually unbiased.

Next, it is clear that the inequality in Eq. (\ref{KD-nonclassicality coherence is upper bounded by purity step 2}) becomes equality when ${\rm Tr}(\varrho^2)=1$, i.e., when the state is pure: $\varrho=\ket{\psi}\bra{\psi}$. Hence, combining with the condition for the equality in Eq. (\ref{the absolute sum of KD nonclassicality and purity step 1}) above, we must have: $|\braket{a|\psi}\braket{\psi|b_*}|=\alpha_{\psi}|\braket{a|b_*}|=\alpha_{\psi}/\sqrt{d}$, for some scalar $\alpha_{\psi}$ which is independent of the index $(a,b_*)$. Following the same argument as for the case of mixed states, this can only be true if $|\braket{a|\psi}|$ and $|\braket{b_*|\psi}|$ are uniform functions independent of $a$ and $b_*$. To guarantee normalization, for $d$-dimensional Hilbert space, we must have $|\braket{a|\psi}|=|\braket{b_*|\psi}|=1/\sqrt{d}$ for all $a,b_*$. This means $\ket{\psi}$ is maximally coherent with respect to both $\{\ket{a}\}$ and $\{\ket{b_*}\}$.   \qed 

As intuitively expected, the upper bound for the KD-nonclassicality coherence of Eq. (\ref{KD-nonclassicality coherence is upper bounded by purity step 1}) decreases as the purity decreases. In particular, when the state is maximally mixed, i.e., $\varrho=\mathbb{I}/d$, the right-hand side of Eq. (\ref{KD-nonclassicality coherence is upper bounded by purity step 1}) is vanishing. Hence, as the amount of the coherence is concerned, the upper bound in Eq. (\ref{KD-nonclassicality coherence is upper bounded by purity step 1}) based on state purity is more informative than that based on measurement uncertainty given in Eq. (\ref{KD negativity is upper bounded by measurement uncertainty 1}). We give numerical examples for the statement of Proposition 3 for the case of single and two qubits in Appendix \ref{Numerical examples of Proposition 3 for single and two qubits}.

\subsection{Lower bound for pure state: optimal guessing probability and trade-off relation\label{Lower bound for pure state: optimal guessing probability and trade-off relation}}

Suppose we perform two measurements described by rank-1 PVM bases $\{\Pi_a\}$ and $\{\Pi_x\}$ of a finite-dimensional Hilbert space $\mathcal{H}$ over the state $\varrho$ on $\mathcal{H}$ with the probabilities of outcomes ${\rm Pr}(a|\varrho)={\rm Tr}(\Pi_a\varrho)$ and ${\rm Pr}(x|\varrho)={\rm Tr}(\Pi_x\varrho)$, respectively. Then, the uncertainties of the measurement outcomes satisfy the Maassen-Uffink uncertainty relation \cite{Massen-Uffink entropic UR}: $H_{\alpha}(\{{\rm Pr}(a|\varrho)\})+H_{\beta}(\{{\rm Pr}(x|\varrho)\})\ge\log 1/c$, where $\alpha,\beta\ge 1/2$ with $1/\alpha+1/\beta=2$. Here, $H_{\alpha}(\{{\rm Pr}(a|\varrho)\})$ is the R\'enyi entropy of index $\alpha$ for the probability of the outcomes of the measurement ${\rm Pr}(a|\varrho)={\rm Tr}(\Pi_a\varrho)$ defined as $H_{\alpha}(\{{\rm Pr}(a|\varrho)\}):=\frac{1}{1-\alpha}\log\sum_a{\rm Pr}(a|\varrho)^{\alpha}$ \cite{Renyi entropy}. Moreover $c$ in the lower bound is defined as 
\begin{eqnarray}
c=\max_{a,x}|\braket{a|x}| 
\label{Massen-Uffink noncommutativity}
\end{eqnarray}
capturing the noncommutativity between the PVM measurement bases $\{\Pi_a\}$ and $\{\Pi_x\}$ which is independent of the state $\varrho$. The lower bound in the Maassen-Uffink uncertainty relation is vanishing when $c=1$ which is the case when the PVM bases $\{\Pi_a\}$ and $\{\Pi_x\}$ share a common element. Moreover, it is maximixed when the PVM bases $\{\Pi_a\}$ and $\{\Pi_x\}$ are complementary corresponding to mutually unbiased orthonormal bases $\{\ket{a}\}$ and $\ket{x}$, so that $c=1/d$. 

Taking $\alpha=1/2$ (or, equivalently, $\beta=\infty$) in the Maassen-Uffink uncertainty relation, we have the following uncertainty relation for min-entropy and max-entropy:
\begin{eqnarray}
H_{\rm max}(\{{\rm Pr}(a|\varrho)\})+H_{\rm min}(\{{\rm Pr}(x|\varrho)\})\ge \log 1/c. 
\label{Maassen-Uffink UR fro max and mix-entropies}
\end{eqnarray} 
Here, the min-entropy in Eq. (\ref{Maassen-Uffink UR fro max and mix-entropies}) is defined as 
\begin{eqnarray}
H_{\rm min}(\{{\rm Pr}(x|\varrho)\}):=H_{\infty}(\{{\rm Pr}(x|\varrho)\})=-\log{\rm Pr}_{\rm guess}(x|\varrho), 
\label{min entropy as uncertainty associated with optimal guessing probability}
\end{eqnarray}
where ${\rm Pr}_{\rm guess}(x|\varrho):=\max_x{\rm Pr}(x|\varrho)$ is the optimal probability of correctly guessing the outcome of measurement described by PVM $\{\Pi_x\}$ hereon called optimal guessing probability. Min-entropy is a central quantity in randomness generation and cryptography. On the other hand, the max-entropy in Eq. (\ref{Maassen-Uffink UR fro max and mix-entropies}) is defined as 
\begin{eqnarray}
H_{\rm max}(\{{\rm Pr}(a|\varrho)\}):=H_{\frac{1}{2}}(\{{\rm Pr}(a|\varrho)\})=2\log\sum_a{\rm Pr}(a|\varrho)^{\frac{1}{2}}. 
\label{max entropy as Renyi 1/2-entropy}
\end{eqnarray}

To connect with the KD-nonclassicality coherence studied in the present work, let us assume that the quantum state is pure, i.e., $\varrho=\ket{\psi}\bra{\psi}$. In this case, noting Proposition 2, and in particular Eq. (\ref{KD-nonclassicality coherence is equal to the measurement uncertainty for pure states}), the max-entropy of Eq. (\ref{max entropy as Renyi 1/2-entropy}) for pure states can be expressed in terms of the KD-nonclassicality coherence as 
\begin{eqnarray}
H_{\rm max}(\{{\rm Pr}(a|\ket{\psi}\bra{\psi})\})=2\log(C_{\rm KD}^{\rm NCl}(\ket{\psi}\bra{\psi};\{\Pi_a\})+1). 
\label{max-entropy in terms of the KD-nonclassicallity coherence}
\end{eqnarray}
Finally, from Eqs. (\ref{Maassen-Uffink UR fro max and mix-entropies}), (\ref{min entropy as uncertainty associated with optimal guessing probability}) and (\ref{max-entropy in terms of the KD-nonclassicallity coherence}), and noting the monotonicity of the log function, the optimal guessing probability in the measurement described by the PVM $\{\Pi_x\}$ is upper bounded by the KD-nonclassicality coherence relative to the orthonormal basis $\{\ket{a}\}$ as 
\begin{eqnarray}
{\rm Pr}_{\rm guess}(x|\ket{\psi}\bra{\psi})\le c(C_{\rm KD}^{\rm NCl}(\ket{\psi}\bra{\psi};\{\Pi_a\})+1)^2. 
\label{KD-nonclassicality coherence as an upper bound for guessing probability}
\end{eqnarray} 
Of particular interest, consider the case when the two PVM bases $\{\Pi_a\}$ and $\{\Pi_x\}$ are complementary, i.e., the associated orthonormal bases $\{\ket{a}\}$ and $\{\ket{x}\}$ are mutually unbiased so that from Eq. (\ref{Massen-Uffink noncommutativity}) we have $c=1/d$. Inserting into Eq. (\ref{KD-nonclassicality coherence as an upper bound for guessing probability}), the optimal guessing probability is upper bounded as 
\begin{eqnarray}
{\rm Pr}_{\rm guess}(x|\ket{\psi}\bra{\psi})\le (C_{\rm KD}^{\rm NCl}(\ket{\psi}\bra{\psi};\{\Pi_a\})+1)^2/d. 
\label{optimal guessing probability for complementary measurement}
\end{eqnarray}
In this case, when the state $\ket{\psi}$ is incoherent relative to $\{\ket{a}\}$ so that $C_{\rm KD}^{\rm NCl}(\ket{\psi}\bra{\psi};\{\Pi_a\})=0$, one has ${\rm Pr}_{\rm guess}(x|\ket{\psi}\bra{\psi})\le 1/d$. Indeed, since $\ket{\psi}$ must be given by one of the elements of $\{\ket{a}\}$, and $\{\ket{x}\}$ is mutually unbiased with $\{\ket{a}\}$, then one must have ${\rm Pr}_{\rm guess}(x|\ket{\psi}\bra{\psi})=1/d$. On the other hand, when the state $\ket{\psi}$ is maximally coherent relative to $\{\ket{a}\}$ so that $C_{\rm KD}^{\rm NCl}(\ket{\psi}\bra{\psi};\{\Pi_a\})=\sum_a\sqrt{{\rm Pr}(a|\ket{\psi}\bra{\psi})}-1=\sqrt{d}-1$ (See Eq. (\ref{KD-nonclassicality coherence is equal to the measurement uncertainty for pure states})), one has ${\rm Pr}_{\rm guess}(x|\ket{\psi}\bra{\psi})\le 1$. In this sense, the KD-nonclassicality coherence of $\ket{\psi}$ relative to the orthonormal basis $\{\ket{a}\}$ is a necessary resource for the optimal probability of correctly guessing the outcomes of measurement described by the PVM basis $\{\Pi_x\}$ maximally noncommuting with the PVM $\{\Pi_a\}$. 

Next, recalling that the R\'enyi entropy is monotonically decreasing as a function of its entropy index, we have $H_{\rm max}(\{{\rm Pr}(a|\varrho)\}):=H_{\frac{1}{2}}(\{{\rm Pr}(a|\varrho)\})\ge H_{\infty}(\{{\rm Pr}(a|\varrho)\}):=H_{\rm min}(\{{\rm Pr}(a|\varrho)\})$. Combining this with Eq. (\ref{Maassen-Uffink UR fro max and mix-entropies}), we thus obtain the following uncertainty relation for the max-entropy: $H_{\rm max}(\{{\rm Pr}(a|\varrho)\})+H_{\rm max}(\{{\rm Pr}(x|\varrho)\})\ge \log 1/c$. For pure state $\varrho=\ket{\psi}\bra{\psi}$, expressing the max-entropy in terms of the KD-nonclassicality coherence as in Eq. (\ref{max-entropy in terms of the KD-nonclassicallity coherence}), we finally obtain the following trade-off relation for the KD-nonclassicality coherences relative to the PVM bases $\{\Pi_a\}$ and $\{\Pi_x\}$:
\begin{eqnarray}
(C_{\rm KD}^{\rm NCl}(\ket{\psi}\bra{\psi};\{\Pi_a\})+1)^2(C_{\rm KD}^{\rm NCl}(\ket{\psi}\bra{\psi};\{\Pi_x\})+1)^2
\nonumber\\ \ge 1/c. 
\label{trade-off relation for the KD-nonclassicality coherence relative to a pair of PVM bases}
\end{eqnarray}
It shows that when $c< 1$, i.e., when the PVM bases $\{\Pi_a\}$ and $\{\Pi_x\}$ are noncommuting, $C_{\rm KD}^{\rm NCl}(\ket{\psi}\bra{\psi};\{\Pi_a\})$ and $C_{\rm KD}^{\rm NCl}(\ket{\psi}\bra{\psi};\{\Pi_x\})$ cannot be both vanishing. As an example, consider the case when $\ket{\psi}$ is given by one of the elements of the orthonormal basis $\{\ket{a}\}$, and the two orthonormal bases $\{\ket{a}\}$ and $\{\ket{x}\}$ are mutually unbiased so that $c=1/d$. In this case, we have $C_{\rm KD}^{\rm NCl}(\ket{\psi}\bra{\psi};\{\Pi_a\})=0$ and $C_{\rm KD}^{\rm NCl}(\ket{\psi}\bra{\psi};\{\Pi_x\})=\sqrt{d}-1$ (as per Proposition 3) so that both sides of Eq. (\ref{trade-off relation for the KD-nonclassicality coherence relative to a pair of PVM bases}) are equal to $d$. Let us emphasize that Eq. (\ref{trade-off relation for the KD-nonclassicality coherence relative to a pair of PVM bases}) applies only for pure state. Otherwise, suppose for example that the state is maximally mixed $\varrho=\mathbb{I}/d$, then we have $C_{\rm KD}^{\rm NCl}(\varrho;\{\Pi_a\})=0$ and $C_{\rm KD}^{\rm NCl}(\varrho;\{\Pi_x\})=0$ for any choice of $\{\Pi_a\}$ and $\{\Pi_x\}$ even when $c< 1$. Hence, for mixed state, we cannot have a trade-off relation similar to Eq. (\ref{trade-off relation for the KD-nonclassicality coherence relative to a pair of PVM bases}) with a state-independent lower bound \cite{Korzekwa quantum-classical decomposition}. 

\section{Operational interpretation: strange weak value and proof of quantum contextuality \label{Experimental estimation, proof of quantum contextuality, and static susceptibility}}

One of the important problems in the quantification of the quantum coherence (or general resource) is to develop a coherence quantifier which can be translated directly to a set of operational procedures in laboratory. Hence, given a quantum system with an unknown quantum state, we wish to directly estimate the coherence of the state relative to an arbitrary orthonormal basis without first estimating the full density matrix via the quantum state tomography. Such an experimental scheme to directly estimate the coherence is important to better understand the operational meaning of quantum coherence. This in turn might also suggest insight into the nature of the nonclassicality associated with quantum coherence.  

Fortunately, the KD-nonclassicality coherence defined in Eq. (\ref{KD-nonclassicality coherence}) can be measured or estimated directly. First, note that the KD quasiprobability associated with a state $\varrho$ over a pair of orthonormal bases $\{\ket{a}\}$ and $\{\ket{b}\}$ can be expressed as 
\begin{eqnarray}
{\rm Pr}_{\rm KD}(a,b|\varrho)=\Pi^{\rm w}_{a}(b|\varrho){\rm Pr}(b|\varrho).
\label{KD quasiprobability as weak value}
\end{eqnarray} 
Here, $\Pi^{\rm w}_{a}(b|\varrho)=\frac{\braket{b|\Pi_a\varrho|b}}{\braket{b|\varrho|b}}$ is just the weak value of the projector $\Pi_a$ with the preselected state $\varrho$ and postselected state $\ket{b}$, and ${\rm Pr}(b|\varrho)=\braket{b|\varrho|b}$ is the probability to get $b$ in the measurement described by the PVM $\{\Pi_b\}$. Different experimental schemes for the estimation of the weak value have been proposed in the literature. The earliest method is by using weak measurement with postselection for the measurement of weak value \cite{Aharonov weak value,Wiseman weak value,Haapasalo generalized weak value,Lundeen complex weak value,Jozsa complex weak value,Aharonov-Daniel book}, based on which the notion of weak value is initially conceived. Since then, different methods without weak measurement have been proposed by different authors \cite{Johansen quantum state from successive projective measurement,Vallone strong measurement to reconstruct quantum wave function,Cohen estimating of weak value with strong measurements,Lostaglio KD quasiprobability and quantum fluctuation,Wagner measuring weak values and KD quasiprobability,Chiribella estimation of weak value,Hernandez-Gomez experimental observation of TBMH negativity}. Let us stress that the real and imaginary parts of the weak value $\Pi^{\rm w}_{a}(b|\varrho)$ can be estimated independently of each other, suggesting that the real and imaginary part of the associated KD quasiprobability ${\rm Pr}_{\rm KD}(a,b|\varrho)$ may capture two distinct notions of quantum statistics.   

Using Eq. (\ref{KD quasiprobability as weak value}), the KD-nonclassicality coherence defined in Eq. (\ref{KD-nonclassicality coherence}), can be expressed as 
\begin{eqnarray}
C_{\rm KD}^{\rm NCl}(\varrho;\{\Pi_a\})=\sup_{\{\ket{b}\}\in\mathcal{B}_{\rm o}(\mathcal{H})}\sum_{a,b}\big|\Pi^{\rm w}_a(b|\varrho)\big|{\rm Pr}(b|\varrho)-1.  
\label{KD-nonclassicality coherence in terms of weak value}
\end{eqnarray}
Eq. (\ref{KD-nonclassicality coherence in terms of weak value}) can then be translated into a hybrid quantum-classical algorithm in the fashion of variational quantum algorithm \cite{Cerezo VQA review} to estimate the KD-nonclassicality coherence of an unknown quantum state directly, without recoursing to the full quantum state tomography. First, let $\ket{b_{\vec{\lambda}}}=U_{\vec{\lambda}}\ket{s}$, where $\{\ket{s}\}$ is a standard orthonormal basis, $U_{\vec{\lambda}}$ is a unitary parameterized by $\vec{\lambda}:=(\lambda_1,\dots,\lambda_M)^{\rm T}$, $M\in\mathbb{N}$, such that varying $\vec{\lambda}$ over all of its range of values leads to scanning the set $\mathcal{B}_{\rm o}(\mathcal{H})$ of all the orthonormal bases $\{\ket{b_{\vec{\lambda}}}\}$ of the Hilbert space $\mathcal{H}$. See e.g. Eqs. (\ref{complete set of bases for two-dimensional Hilbert space}) and (\ref{product bases of two qubit}) for the parameterization of the orthonormal (product) bases for the case of one and two qubits, respectively. Let us assume that such a parameterized unitary $U_{\vec{\lambda}}$ which transforms a standard orthonormal basis into all the orthonormal bases of the Hilbert space can be implemented in a quantum circuit. The algorithm for the estimation of the KD-nonclassicality coherence then goes as follows. \\
\begin{enumerate}[font={\bfseries},label={S\arabic*}]
\item Estimate the real and imaginary parts of the weak value $\Pi^{\rm w}_a(b_{\vec{\lambda}}|\varrho)$ using one of the schemes proposed in the literatures, with the input: the unknown quantum state $\varrho$, the element of the incoherent orthonormal basis $\{\ket{a}\}$, and the element of the second orthonormal basis $\{\ket{b_{\vec{\lambda}}}\}$. \\
\item Take the average of the modulus weak value $\Pi^{\rm w}_a(b_{\vec{\lambda}}|\varrho)$ over the probability ${\rm Pr}(b_{\vec{\lambda}}|\varrho)$ and sum over all index $a$ following the prescription of Eq. (\ref{KD-nonclassicality coherence in terms of weak value}). \\
\item Vary the parameters $\vec{\lambda}$ of the unitary $U_{\vec{\lambda}}$ that prepares the second basis $\{\ket{b_{\vec{\lambda}}}\}\in\mathcal{B}_{\rm o}(\mathcal{H})$. 
\item Repeat S1-S3 to obtain a converging supremum value.  
\end{enumerate}

The operational interpretation of the KD-nonclassicality coherence in terms of the weak value measurement leads to the following conceptual connection between the KD-nonclassicality coherence and the quantum contextuality \cite{Spekkens quantum contextuality}. First, one can see in Eq. (\ref{KD quasiprobability as weak value})  that the nonreal values and the negative values of the real part of the KD quasiprobability ${\rm Pr}_{\rm KD}(a,b|\varrho)$ appears in the nonreal values and negative values of the real part of the weak value $\Pi^{\rm w}_{a}(b|\varrho)$. Such nonreal values and negative values of the real part of the weak value of a projector are called strange weak values. It has been shown in Refs.  \cite{Pusey negative TBMH quasiprobability and contextuality,Kunjwal contextuality of non-real weak value,Lostaglio contextuality in quantum linear response} that when the weak value $\Pi^{\rm w}_{a}(b|\varrho)$ is strange, its estimation based on weak measurement with postselection cannot be simulated using noncontextual hidden variable model. Hence, the nonreality or the negativity of the KD quasiprobability can be used to prove quantum contextuality. 

The above observation implies that a nonvanishing KD-nonclassicality coherence of a state $\varrho$ relative to an orthonormal basis $\{\Pi_a\}$ is sufficient and necessary for the proof of the quantum contextuality via the estimation of the weak value $\Pi_a^{\rm w}(b|\varrho)$ using weak measurement with postselection. As for the sufficiency, first, suppose that the KD-nonclassicality coherence of a state $\varrho$ relative to an orthonormal basis $\{\ket{a}\}$ is nonvanishing. Then, there must be at least a single element of the rank-1 PVM $\{\Pi_a\}$ such that its weak value $\Pi_a^{\rm w}(b|\varrho)$ is nonreal or its real part is negative. This can then be used to prove quantum contextuality via weak measurement with postselection as shown in Refs. \cite{Pusey negative TBMH quasiprobability and contextuality,Kunjwal contextuality of non-real weak value,Lostaglio contextuality in quantum linear response}. Conversely, suppose that the weak value $\Pi_a^{\rm w}(\phi|\varrho)$ for some $\ket{\phi}$ is strange, i.e., nonreal and/or its real part is negative, so that it can be used to prove quantum contextuality. Then, one can choose an orthonormal basis $\{\ket{b}\}$ such that one of its element is given by $\ket{\phi}$. In this case, the KD-nonclassicality relative to the orthonormal bases $\{\Pi_a\}$ and $\{\Pi_b\}$ defined in Eq. (\ref{KD nonclassicality}) is nonvanishing: ${\rm NCl}(\{{\rm Pr}_{\rm KD}(a,b|\varrho\})=\sum_{a,b}|\Pi^{\rm w}_a(b|\varrho)|\braket{b|\varrho|b}-1>0$. This further implies that the KD-nonclassicality coherence defined in Eq. (\ref{KD-nonclassicality coherence}) must also be nonvanishing: $C_{\rm KD}^{\rm NCl}(\varrho;\{\Pi_a\})=\sup_{\{\ket{b}\}\in\mathcal{B}_{\rm o}(\mathcal{H})}{\rm NCl}(\{{\rm Pr}_{\rm KD}(a,b|\varrho\})>0$. 

\section{Concluding Remarks\label{Summary and Remarks}}  

The KD-nonclassicality coherence proposed in this work defined in Eq. (\ref{KD-nonclassicality coherence}) complements the KD-nonreality coherence based on solely the total sum of the absolute nonreal part of the KD quasiprobability defined in Eq. (\ref{KD-nonreality coherence}) proposed in Ref. \cite{Agung KD-nonreality coherence}. In particular, by using KD nonclassicality which captures simultaneously the nonreality and negativity of the KD quasiprobability, we have been able to connect quantum coherence with the measurement uncertainty quantified by the Tsallis $\frac{1}{2}$-entropy. For pure states, this in turn allows us to combine it with the Maassen-Uffink uncertainty relation for min-entropy and max-entropy, to use the KD-nonclassicality coherence to upper bound the optimal guessing probability, and to derive a trade-off relation for the KD-nonclassicality coherences with Massen-Uffink-like state-independent lower bound. Furthermore, using the KD nonclassicality for the quantification of the coherence also reveals a connection between quantum coherence and proof of quantum contextuality, in similar fashion as the connection between entanglement and proof of quantum nonlocality. Our results thus strengthen the quantitative link between the quantumness manifests in the nonclassical values of KD quasiprobability and the associated strange weak value to the nonclassicality captured by the concept of quantum state subjected to measurement in the forms of asymmetry \cite{Agung translational asymmetry from nonreal weak value,Agung trace-norm asymmetry} and general quantum correlation \cite{Agung KD general quantum correlation}. These results motivate the search for a quantitative connection between KD nonclassicality and strange weak value to quantum entanglement.   

\begin{acknowledgments}  
This work is partly funded by the Institute for Research and Community Service, Bandung Institute of Technology with the grant number: 2971/IT1.B07.1/TA.00/2021. It is also in part supported by the Indonesia Ministry of Research, Technology, and Higher Education with the grant number: 187/E5/PG.02.00.PT/2022. 
\end{acknowledgments}

\appendix

\section{Proofs of properties (i)-(vii)\label{Proofs of Properties (i)-(vi)}}

Here, we give the proofs of the properties (i)-(vii) stated in Section \ref{Quantum coherence via the KD nonclassicality} of the main text. 
\\
{\bf Property (i)} {\it Faithfulness}. \\
{\bf Proof}. Suppose first that the quantum state $\varrho$ is incoherent with respect to an orthonormal basis $\{\ket{a}\}$. It can thus be decomposed as $\varrho=\sum_ap_a\Pi_a$, where $p_a\ge 0$, $\sum_ap_a=1$, so that $[\Pi_a,\varrho]=0$ for all $a$. This implies ${\rm Im}{\rm Pr}_{\rm KD}(a,b|\varrho)={\rm Tr}(\Pi_b[\Pi_a,\varrho])/2i=0$ for all $a$ and all the orthnormal bases $\{\ket{b}\}\in\mathcal{B}_{\rm o}(\mathcal{H})$ of the Hilbert space $\mathcal{H}$. Hence, the KD quasiprobability is real. Moreover, in this case, we have ${\rm Re}{\rm Pr}_{\rm KD}(a,b|\varrho)=|\braket{a|b}|^2p_a\ge 0$ for all $a$ and all $\{\ket{b}\}\in\mathcal{B}_{\rm o}(\mathcal{H})$, so that the real part is nonnegative. The above facts that the KD quasiprobability is real and nonnegative imply $|{\rm Pr}_{\rm KD}(a,b|\varrho)|={\rm Pr}_{\rm KD}(a,b|\varrho)$ for all $a$ and and all $\{\ket{b}\}\in\mathcal{B}_{\rm o}(\mathcal{H})$, so that we have ${\rm NCl}(\{{\rm Pr}_{\rm KD}(a,b|\varrho)\})=\sum_{a,b}|{\rm Pr}_{\rm KD}(a,b|\varrho)|-1=\sum_{a,b}{\rm Pr}_{\rm KD}(a,b|\varrho)-1=0$, for all $\{\ket{b}\}\in\mathcal{B}_{\rm o}(\mathcal{H})$, where the last equality follows from the normalization of the KD quasiprobability. We thus have, by definition, $C_{\rm KD}^{\rm NCl}(\varrho;\{\Pi_a\})=\sup_{\{\ket{b}\}\in\mathcal{B}_{\rm o}(\mathcal{H})}{\rm NCl}(\{{\rm Pr}_{\rm KD}(a,b|\varrho)\})=0$. 

Conversely, suppose $C_{\rm KD}^{\rm NCl}(\varrho;\{\Pi_a\})=0$. Then by definition, we must have $\sum_{a,b}|\braket{b|\Pi_a\varrho|b}|=1$ for all $\{\ket{b}\}\in\mathcal{B}_{\rm o}(\mathcal{H})$. Due to the normalization of KD quasiprobability, this can only be true when $|\braket{b|\Pi_a\varrho|b}|=\braket{b|\Pi_a\varrho|b}$ for all $a$ and all $\{\ket{b}\}\in\mathcal{B}_{\rm o}(\mathcal{H})$. Namely, for all $a$ and all $\{\ket{b}\}\in\mathcal{B}_{\rm o}(\mathcal{H})$, we must have ${\rm Im}\braket{b|\Pi_a\varrho|b}=0$, i.e., the KD quasiprobability must be real, and ${\rm Re}\braket{b|\Pi_a\varrho|b}\ge0$, i.e, the real part of the KD quasiprobability must be nonnegative. From the reality of the KD quasiprobability alone we have $0={\rm Im}\braket{b|\Pi_a\varrho|b}=\frac{1}{2i}\braket{b|[\Pi_a,\varrho]|b}$ for all $a$ and all $\{\ket{b}\}\in\mathcal{B}_{\rm o}(\mathcal{H})$. Hence, we have $[\Pi_a,\varrho]=0$ for all $a$. It means that $\{\Pi_a\}$ is the set of eigenprojectors of $\varrho$ to have $\varrho=\sum_ap_a\Pi_a$, with $p_a\ge 0$, $\sum_ap_a=1$, i.e., $\varrho$ is incoherent relative to the orthonormal basis $\{\ket{a}\}$. \qed
\\
{\bf Property (ii)} {\it Unitarily covariant}.\\
{\bf Proof}. It can be established from the definition as follows
\begin{eqnarray}
&&C_{\rm KD}^{\rm NCl}( U\varrho U^{\dagger};\{ U\Pi_a U^{\dagger}\})\nonumber\\
&=&\sup_{\{\ket{b}\}\in\mathcal{B}_{\rm o}(\mathcal{H})}\sum_a\sum_b\big|\braket{b| U\Pi_a U^{\dagger} U\varrho U^{\dagger}|b}\big|-1\nonumber\\
&=&\sup_{\{\ket{b_U}\}\in\mathcal{B}_{\rm o}(\mathcal{H})}\sum_a\sum_{b_U}\big|\braket{b_U|\Pi_a\varrho|b_U}\big|-1\nonumber\\
&=&C_{\rm KD}^{\rm NCl}(\varrho;\{\Pi_a\}), 
\label{proof of the unitary covariant property}
\end{eqnarray}
where we have taken into account the fact that the unitary operator $U$ leads to transformation between orthonormal bases $\{\ket{b_U}\}=\{U^{\dagger}\ket{b}\}$ of the same Hilbert space $\mathcal{H}$, so that the set of the new orthonormal bases $\{\ket{b_U}\}$ is the same as that of the old orthonormal bases $\{\ket{b}\}$ given by $\mathcal{B}_{\rm o}(\mathcal{H})$. This implies $\sup_{\{\ket{b_U}\}\in\mathcal{B}_{\rm o}(\mathcal{H})}(\cdot)=\sup_{\{\ket{b}\}\in\mathcal{B}_{\rm o}(\mathcal{H})}(\cdot)$.  \qed
\\
{\bf Property (iii)} {\it Convexity}. \\
{\bf Proof}. This is a direct implication of the triangle inequality, and the fact that $p_k\ge 0$ and $\sum_kp_k=1$, i.e., \begin{eqnarray}
&&C_{\rm KD}^{\rm Ncl}\big(\sum_kp_k\varrho_k;\{\Pi_a\}\big)\nonumber\\
&=&\sup_{\{\ket{b}\}\in\mathcal{B}_{\rm o}(\mathcal{H})}\sum_a\sum_b\big|\braket{b|\Pi_a\sum_kp_k\varrho_k|b}\big|-1\nonumber\\
&\le&\sum_kp_k\sup_{\{\ket{b}\}\in\mathcal{B}_{\rm o}(\mathcal{H})}\sum_a\sum_b\big|\braket{b|\Pi_a\varrho_k|b}\big|-1\nonumber\\
&=&\sum_kp_k\Big(\sup_{\{\ket{b}\}\in\mathcal{B}_{\rm o}(\mathcal{H})}\sum_a\sum_b\big|\braket{b|\Pi_a\varrho_k|b}\big|-1\Big)\nonumber\\
&=&\sum_kp_kC_{\rm KD}^{\rm NCl}(\varrho_k;\{\Pi_a\}).
\end{eqnarray} \qed
\\
{\bf Property (iv)} {\it Nonincreasing under partial trace}.\\
{\bf Proof}. This can be shown straightforwardly as
\begin{eqnarray}
&&C_{\rm KD}^{\rm NCl}(\varrho_{12};\{\Pi_{a_1}\otimes\mathbb{I}_2\})\nonumber\\
&:=&\sup_{\{\ket{b_1,b_2}\}\in\mathcal{B}_{\rm op}(\mathcal{H}_{12})}\sum_{a_1}\sum_{b_1,b_2}\big|\sum_{a_2}{\rm Pr}_{\rm KD}(a_1,a_2,b_1,b_2|\varrho_{12})\big|-1\nonumber\\
&=&\sup_{\{\ket{b_1,b_2}\}\in\mathcal{B}_{\rm op}(\mathcal{H}_{12})}\sum_{a_1}\sum_{b_1,b_2}\big|\braket{b_1,b_2|(\Pi_{a_1}\otimes\mathbb{I}_2)\varrho_{12}|b_1,b_2}\big|-1\nonumber\\
&\ge&\sup_{\{\ket{b_1,b_2}\}\in\mathcal{B}_{\rm op}(\mathcal{H}_{12})}\sum_{a_1}\sum_{b_1}\big|\sum_{b_2}\braket{b_1,b_2|(\Pi_{a_1}\otimes\mathbb{I}_2)\varrho_{12}|b_1,b_2}\big|-1\nonumber\\
&=&\sup_{\{\ket{b_1}\}\in\mathcal{B}_{\rm o}(\mathcal{H}_1)}\sum_{a_1}\sum_{b_1}\big|\braket{b_1|\Pi_{a_1}\varrho_1|b_1}\big|-1\nonumber\\
&=& C_{\rm KD}^{\rm NCl}(\varrho_1;\{\Pi_{a_1}\}).
\label{proof of partial trace}
\end{eqnarray}
Here, $\mathcal{B}_{\rm op}(\mathcal{H}_{12})$ is the set of all orthonormal product bases of the Hilbert space $\mathcal{H}_{12}=\mathcal{H}_1\otimes\mathcal{H}_2$ of the bipartite system $12$, $\varrho_{12}$ is a bipartite state on $\mathcal{H}_{12}$, and $\varrho_1=\sum_{b_2}\braket{b_2|\varrho_{12}|b_2}={\rm Tr}_2(\varrho_{12})$ is the reduced state of system 1 on $\mathcal{H}_1$. One can see that equality in Eq. (\ref{proof of partial trace}) is obtained when there is no quantum and classical correlation in the quantum state, i.e., $\varrho_{12}=\varrho_1\otimes\varrho_2$, by virtue of the fact that $\braket{b_2|\varrho_2|b_2}$ is real and nonnegative for all $b_2$, and the normalization $\sum_{b_2}\braket{b_2|\varrho_2|b_2}=1$. \qed
\\
{\bf Property (v)} {\it Nonincreasing under decoherence operation}.\\
{\bf Proof}. Using the property (iii) of convexity, we have for $0\le p\le 1$,
\begin{eqnarray}
&&C_{\rm KD}^{\rm Ncl}\big(p\varrho+(1-p)\mathcal{D}(\varrho;\{\Pi_a\});\{\Pi_a\}\big)\nonumber\\
&\le&pC_{\rm KD}^{\rm Ncl}\big(\varrho;\{\Pi_a\})+(1-p)C_{\rm KD}^{\rm NCl}(\mathcal{D}(\varrho;\{\Pi_a\});\{\Pi_a\}\big)\nonumber\\
&=&pC_{\rm KD}^{\rm NCl}(\varrho;\{\Pi_a\})\le C_{\rm KD}^{\rm NCl}(\varrho;\{\Pi_a\}),
\end{eqnarray}
where, to get the third line, we have noted the fact that $[\mathcal{D}(\varrho;\{\Pi_a\}),\Pi_{a}]=0$ for all $a$ so that $C_{\rm KD}^{\rm Ncl}\big(\mathcal{D}(\varrho;\{\Pi_a\});\{\Pi_a\}\big)=0$ by virtue of property (i).  \qed
\\
{\bf Property (vi)} {\it Nonincreasing under coarsegraining}.\\
{\bf Proof}. This follows directly from the definition as
\begin{eqnarray}
&&C_{\rm KD}^{\rm NCl}(\varrho;\{\Pi_A\})\nonumber\\
&:=&\sup_{\{\ket{b}\}\in\mathcal{B}_{\rm o}(\mathcal{H})}\sum_{A,b}|{\rm Pr}_{\rm KD}(A,b))|-1\nonumber\\
&=&\sup_{\{\ket{b}\}\in\mathcal{B}_{\rm o}(\mathcal{H})}\sum_{A,b}\big|\sum_{a\in A}{\rm Tr}(\Pi_b\Pi_a\varrho))\big|-1\nonumber\\
&\le&\sup_{\{\ket{b}\}\in\mathcal{B}_{\rm o}(\mathcal{H})}\sum_{a,b}|{\rm Tr}(\Pi_b\Pi_a\varrho)|-1\nonumber\\
&=&C_{\rm KD}^{\rm NCl}(\varrho;\{\Pi_a\}).
\end{eqnarray}
\qed
\\
{\bf Property (vii)} {\it Nonincreasing under a controlloed incoherent permutation of the element of the incoherent basis}. \\
Before proceeding with a proof, let us first justify the plausibility of the conditions for the property (vii) mentioned in the main text. Consider a class of completely positive trace-preserving linear maps or quantum channel which can be expressed in terms of free dilation as follows: 
\begin{eqnarray}
\Phi_{\rm CIP}(\varrho_S)={\rm Tr}_E\big(U_{SE}(\varrho_S\otimes\varrho_E)U_{SE}^{\dagger}\big), 
\label{trace-preserving QO}
\end{eqnarray}
where $\varrho_S$ is the state of the (principal) system, $\varrho_E$ is the state of an ancilla (auxiliary system), $U_{SE}$ is a unitary which couples the system and the ancilla, and the partial trace is taken over the ancilla. Moreover, by free dilation we mean that the following conditions are satisfied. First, the state of the ancilla $\varrho_E$ is incoherent relative to an a priori fixed basis $\{\ket{e}_E\}$ of the ancillary system. And second, the action of the coupling unitary takes the form: 
\begin{eqnarray}
U_{SE}\ket{a}_S\ket{e}_E=\ket{\mu_e(a)}_S\ket{e}_E,
\label{free entangling unitaries}
\end{eqnarray}
where $\ket{a}_S$ is the element of the incoherent basis of the principal system and $\{\mu_{e}(a)\}$ is a set of permutations of the elements of the incoherent basis of the principal system depending on the element of the incoherent basis of the ancilla. $U_{SE}$ is therefore a control incoherent permutation transformation. For example, if both the system $S$ and the ancilla $E$ are qubit (two-dimensional system), and their incoherent bases are given by the computational basis $\{\ket{0},\ket{1}\}$, and assuming $\mu_e(a)=a+e\mod{2}$, then $U_{SE}$ is just the CNOT gate. $U_{SE}$ thus generalizes the free permutation unitaries in the resource theory of coherence by making it dependent on the state of the ancilla. 

We can then state the following condition mentioned in the Property (vii).\\
{\bf Proposition 4}. The completely positive trace-preserving quantum operation $\Phi_{\rm CIP}(\cdot)$ defined in Eq. (\ref{trace-preserving QO}) is incoherence preserving, namely it maps an incoherent state onto an incoherent state. \\
{\bf Proof}. Consider an arbitrary incoherent state of the principal system: $\varrho_S=\sum_ap_a\ket{a}\bra{a}$, where $\{p_a\}$, $p_a\ge 0$, $\sum_ap_a=1$, are mixing probabilities. Further, since the state of the ancilla $\varrho_E$ is incoherent with respect to the ancillary basis $\{\ket{e}\}$, it can be decomposed as $\varrho_E=\sum_er_e\ket{e}\bra{e}$, where $\{r_e\}$, $r_e\ge 0$, $\sum_er_e=1$, are probabilities. Then one has, upon employing Eq. (\ref{free entangling unitaries}), 
\begin{eqnarray}
&&\Phi_{\rm CIP}(\varrho_S)\nonumber\\
&=&{\rm Tr}_E\big(U_{SE}(\varrho_S\otimes\varrho_E)U_{SE}^{\dagger}\big)\nonumber\\
&=&\sum_{a,e,e'}p_ar_e{\rm Tr}_E\big(U_{SE}(\ket{a}\bra{a}\otimes\ket{e}\bra{e})U_{SE}^{\dagger}\big)\nonumber\\
&=&\sum_{a,e}p_ar_e{\rm Tr}_E\big(\ket{\mu_e(a)}\bra{\mu_e(a)}\otimes\ket{e}\bra{e}\big)\nonumber\\
\label{stochastic permutation is incoherence-preserving step 1}
&=&\sum_er_e\sum_ap_a\ket{\mu_e(a)}\bra{\mu_e(a)}. 
\end{eqnarray}
$\Phi_{\rm CIP}(\varrho_S)$ is thus a statistical mixture of the elements of the incoherent basis, hence it is again an incoherent state. It can also be seen in Eq. (\ref{stochastic permutation is incoherence-preserving step 1}) that the quantum operation $\Phi_{\rm CIP}(\varrho_S)$ maps the incoherent state as: $\sum_ap_a\ket{a}\bra{a}\mapsto\sum_ap_a\ket{\mu_e(a)}\bra{\mu_e(a)}$, by permuting each elements as $a\mapsto\mu_e(a)$ with a probability $r_e=\braket{e|\varrho_E|e}$, and then delete the records on $e$. \qed   

With the above clarification, the proof of Property (vii) then proceeds as follows. \\  
{\bf Proof}. First, using the Stinespring's dilation we have 
\begin{eqnarray}
&&C_{\rm KD}^{\rm NCl}(\Phi_{\rm CIP}(\varrho_S);\{\Pi_a\})+1\nonumber\\
&=&\sup_{\{\ket{b}\}\in\mathcal{B}_{\rm o}(\mathcal{H})}\sum_{a,b}\big|\braket{b|\Pi_a{\rm Tr}_E\big(U_{SE}(\varrho_S\otimes\varrho_E)U_{SE}^{\dagger}\big)|b}\big|\nonumber\\
\label{proof of property 8 step 1}
&\le&\sup_{\{\ket{b}\}\in\mathcal{B}_{\rm o}(\mathcal{H})}\sum_{a,b,e}\big|\braket{b,e|(\Pi_a\otimes\mathbb{I}_E)U_{SE}(\varrho_S\otimes\varrho_E)U_{SE}^{\dagger}|b,e}\big|\\
\label{proof of property 8 step 1.5}
&=&\sup_{\{\ket{b}\}\in\mathcal{B}_{\rm o}(\mathcal{H})}\sum_{a,b,e}\big|\braket{b,e|U_{SE}U_{SE}^{\dagger}(\Pi_a\otimes\Pi_e)U_{SE}(\varrho_S\otimes\varrho_E)U_{SE}^{\dagger}|b,e}\big|\\
\label{proof of property 8 step 2}
&=&\sup_{\{\ket{b}\}\in\mathcal{B}_{\rm o}(\mathcal{H})}\sum_{\mu_e^{-1}(a),b,e}\big|\braket{b,e|U_{SE}(\Pi_{\mu^{-1}_e(a)}\otimes\Pi_e)(\varrho_S\otimes\varrho_E)U_{SE}^{\dagger}|b,e}\big|. 
\end{eqnarray}
Here, the basis of the ancilla inserted in Eq. (\ref{proof of property 8 step 1}) is arbitrary and we have chosen the incoherent basis $\{\ket{e}\}$, in Eq. (\ref{proof of property 8 step 1.5}) we have inserted the identity $U_{SE}U_{SE}^{\dagger}=\mathbb{I}_{SE}$, and to get Eq. (\ref{proof of property 8 step 2}) we have used $U_{SE}^{\dagger}\ket{a}\otimes\ket{e}=\ket{\mu_e^{-1}(a)}\otimes\ket{e}$ which follows from the definition of the controlled incoherent permutation unitary in Eq. (\ref{free entangling unitaries}), where $\mu_e^{-1}(a)$ is the inverse permutation of $\mu(a)$. Next, expanding $\ket{b}=\sum_a\braket{a|b}\ket{a}$, and again using $U_{SE}^{\dagger}\ket{a}\otimes\ket{e}=\ket{\mu_e^{-1}(a)}\otimes\ket{e}$, we get 
\begin{eqnarray}
U_{SE}^{\dagger}\ket{b,e}=\sum_a\braket{a|b}\ket{\mu_e^{-1}(a)}\ket{e}=\ket{\nu_e(b)}\ket{e}, 
\label{proof of property 8 step 3}
\end{eqnarray}
where $\ket{\nu_e(b)}:=\sum_a\braket{a|b}\ket{\mu_e^{-1}(a)}$. Moreover, one can check that for each $e$, the set $\{\ket{\nu_e(b)}\}$ comprises an orthonormal basis of the Hilbert space, i.e., $\braket{\nu_e(b)|\nu_e(b')}=\delta_{bb'}$ and $\sum_b\ket{\nu_e(b)}\bra{\nu_e(b)}=\mathbb{I}$. Inserting Eq. (\ref{proof of property 8 step 3}) into Eq. (\ref{proof of property 8 step 2}), we thus obtain
\begin{eqnarray}
&&C_{\rm KD}^{\rm NCl}(\Phi_{\rm CIP}(\varrho_S);\{\Pi_a\})+1\nonumber\\
&\le&\sup_{\{\ket{\nu_e(b)}\}\in\mathcal{B}_{\rm o}(\mathcal{H})}\sum_{\mu_e^{-1}(a),\nu_e(b),e}\big|\braket{\nu_e(b),e|(\Pi_{\mu^{-1}_e(a)}\otimes\Pi_e)(\varrho_S\otimes\varrho_E)|\nu_e(b),e}\big|\nonumber\\
\label{monoton step 1}
&=&\sup_{\{\ket{b}\}\in\mathcal{B}_{\rm o}(\mathcal{H})}\sum_{a,b,e}|\braket{b|\Pi_a\varrho_S|b}|\braket{e|\varrho_E|e}\\
\label{monoton step 2}
&=&\sum_e\big(C_{\rm KD}^{\rm NCl}(\varrho_S;\{\Pi_a\})+1\big)\braket{e|\varrho_E|e}\\
\label{monoton step 3}
&=&C_{\rm KD}^{\rm NCl}(\varrho_S;\{\Pi_a\})+1. 
\end{eqnarray}
Here, in Eq. (\ref{monoton step 1}) we have made relabelling of the indexes of the bases of the principal system and noted the fact that $\braket{e|\varrho_E|e}\ge 0$, to get Eq. (\ref{monoton step 2}) we have used the definition of the KD-nonclassicality coherence, and Eq. (\ref{monoton step 3}) is due to the normalization $\sum_e\braket{e|\varrho_E|e}=1$. \qed

\section{Proof of Proposition 1\label{Proof of Proposition 1}}

One first has, from the definition of the KD-nonclassicality coherence in Eq. (\ref{KD-nonclassicality coherence}), 
\begin{eqnarray}
&&C_{\rm KD}^{\rm NCl}(\varrho;\{\Pi_a\})\nonumber\\
&=&\sup_{\{\ket{b}\}\in\mathcal{B}_{\rm o}(\mathcal{H})}\sum_{a,b}\big|\sum_{a'}\braket{b|a}\braket{a|\varrho|a'}\braket{a'|b}\big|-1\nonumber\\
&\le&\sup_{\{\ket{b}\}\in\mathcal{B}_{\rm o}(\mathcal{H})}\sum_{a,a',b}\big|\braket{a|\varrho|a'}\braket{b|a}\braket{a'|b}\big|-1\nonumber\\
&=&\sum_{a,a'}\big|\braket{a|\varrho|a'}|\sum_{b_*}|\braket{b_*|a}\braket{a'|b_*}\big|-1, 
\label{KD-nonclassicality coherence is upper bounded by l1 coherence step 0}
\end{eqnarray}
where $\{\ket{b_*}\}\in\mathcal{B}_{\rm o}(\mathcal{H})$ is a second basis which achieves the supremum. On the other hand, using the Cauchy-Schwartz inequality, we have $\sum_{b_*}|\braket{b_*|a}\braket{a'|b_*}|\le(\sum_{b_*}|\braket{b_*|a}|^2\sum_{b_*'}|\braket{a'|b'_*}|^2)^{1/2}=1$, where we have used the completeness relation $\sum_{b_*}\ket{b_*}\bra{b_*}=\mathbb{I}$. Moreover, the equality (i.e., the maximum) is obtained when the pair of bases $\{\ket{a}\}$ and $\{\ket{b_*}\}$ are mutually unbiased so that $|\braket{a|b_*}|=1/\sqrt{d}$ for all ${(a,b_*)}$. Inserting this into Eq. (\ref{KD-nonclassicality coherence is upper bounded by l1 coherence step 0}), we finally obtain
\begin{eqnarray}
C_{\rm KD}^{\rm NCl}(\varrho;\{\Pi_a\})&\le&\sum_{a,a'}\big|\braket{a|\varrho|a'}|-1\nonumber\\
&=&\sum_{a\neq a'}\big|\braket{a|\varrho|a'}|\nonumber\\
&=&C_{l_1}(\varrho;\{\ket{a}\}),
\label{KD-nonclassicality coherence is upper bounded by l1 coherence}
\end{eqnarray} 
where we have used the normalization $\sum_a\braket{a|\varrho|a}=1$. \qed

\section{Numerical examples of Proposition 3 for single and two qubits\label{Numerical examples of Proposition 3 for single and two qubits}}

We give a few numerical examples for single and two qubits confirming the statement of Proposition 3. 
\\
{\bf Example 1: a single qubit}. Consider a single qubit with a density matrix in the basis $\{\ket{a}\}=\{\ket{0},\ket{1}\}$ that is given by Eq. (\ref{density matrix for a single qubit in Bloch sphere}) with $\varphi=0$. As shown in Fig. 1, in this case, the upper bound of the KD-nonclassicality coherence relative to the incoherent basis $\{\ket{0},\ket{1}\}$ of Eq. (\ref{KD-nonclassicality coherence is upper bounded by purity step 1}), i.e., $\sqrt{2{\rm Tr}(\varrho^2)}-1=\sqrt{(1+r^2)}-1$, for different values of $r$ are attained when $\theta=\pi/2$. Moreover, we also found numerically that the second basis $\{\ket{b_*}\}$ which achieves the supremum in Eq. (\ref{KD-nonclassicality coherence}) is mutually unbiased with the incoherent basis $\{\ket{0},\ket{1}\}$ in accord with Proposition 3. For instance, consider the case when $r=1$ and $\varphi=0$ so that the quantum state reads $\ket{\psi}=\cos\frac{\theta}{2}\ket{0}+\sin\frac{\theta}{2}\ket{1}$. Then, the maximum of the KD-nonclassicality coherence, which is equal to $\sqrt{2}-1$, is obtained when $\theta=\pi/2$ (see Fig. 1), i.e., when the state is maximally coherent $\ket{\psi}=\frac{1}{\sqrt{2}}(\ket{0}+\ket{1})$. Moreover, in this case, the second basis $\{\ket{b_*}\}$ which solves the optimization in Eq. (\ref{KD-nonclassicality coherence}) is given by $\{\ket{b_*}\}=\{\ket{y_+},\ket{y_-}\}$, where $\ket{y_{\pm}}=\frac{1}{\sqrt{2}}(\ket{0}\pm i\ket{1})$, which is mutually unbiased with the incoherent basis $\{\ket{a}\}=\{\ket{0},\ket{1}\}$. Notice that the state $\ket{\psi}=\frac{1}{\sqrt{2}}(\ket{0}+\ket{1})$ is also maximally coherent with respect to the optimal second basis $\{\ket{b_*}\}=\{\ket{y_+},\ket{y_-}\}$, in accord with Proposition 3. One can further check that for $\ket{\psi}=\frac{1}{\sqrt{2}}(\ket{0}+\ket{1})$, the associated KD quasiprobability over the pair of orthonormal bases $\{\ket{a}\}=\{\ket{0},\ket{1}\}$ and $\{\ket{b_*}\}=\{\ket{y_+},\ket{y_-}\}$, has the form, in matrix expression with the incoherent basis $\{\ket{a}\}=\{\ket{0},\ket{1}\}$ taking the row index and the second optimal basis $\ket{b_*}=\{\ket{y_+},\ket{y_-}\}$ taking the column index:
\begin{eqnarray}
\{{\rm Pr}_{\rm KD}(\varrho)\}_{ab_*}= 
\begin{pmatrix} 
\frac{1}{4}+\frac{1}{4}i & \frac{1}{4}-\frac{1}{4}i& \\
\frac{1}{4}-\frac{1}{4}i & \frac{1}{4}+\frac{1}{4}i&
\label{KD quasiprobability for maximally coherent state}
\end{pmatrix}.
\end{eqnarray}
Notice that the real part of the above KD quasiprobability are all nonnegative. Hence, the coherence is captured entirely by the nonreality of the KD quasiprobability.   
\\
{\bf Example 2: maximally coherent two-qubit state}. Consider two-qubit system, i.e., $d=4$, with a maximally coherent state given by $\ket{\psi}=(\ket{00}-i\ket{01}-i\ket{10}+i\ket{11})/2$.  Let us compute the KD-nonclassicality coherence of the state relative to the incoherent orthonormal product basis $\{\ket{a}\}=\{\ket{00},\ket{01},\ket{10},\ket{11}\}$. According to the Remark following Definition 2, the second basis is given also by a product basis $\{\ket{b}\}=\{\ket{b_{1+},b_{2+}},\ket{b_{1+},b_{2-}},\ket{b_{1-},b_{2+}},\ket{b_{1-},b_{2-}}\}$. The second basis can be expressed using the Bloch sphere parameterization as
\begin{eqnarray}
\ket{b_{k+}(\alpha_k,\beta_k)}&:=&\cos\frac{\alpha_k}{2}\ket{0}_k+\sin\frac{\alpha_k}{2}e^{i\beta_k}\ket{1}_k;\nonumber\\
\ket{b_{k-}(\alpha_k,\beta_k)}&:=&\sin\frac{\alpha_k}{2}\ket{0}_k-\cos\frac{\alpha_k}{2}e^{i\beta_k}\ket{1}_k, 
\label{product bases of two qubit}
\end{eqnarray}
$\alpha_k\in[0,\pi]$, $\beta_k\in[0,2\pi)$, $k=1,2$. Then, one obtains, numerically that the KD-nonclassicality coherence is maximized, i.e., $C_{\rm KD}^{\rm NCl}(\ket{\psi}\bra{\psi};\{\ket{00}\bra{00},\ket{01}\bra{01},\ket{10}\bra{10},\ket{11}\bra{11}\})=\sqrt{4}-1=1$, saturating the inequality (\ref{KD-nonclassicality coherence is upper bounded by purity step 2}), in accord with Proposition 3. One can also check this analytically using Eq. (\ref{KD-nonclassicality coherence is equal to the measurement uncertainty for pure states}) for pure state. The second product basis $\{\ket{b_*}\}$ which attains the supremum in Eq. (\ref{KD-nonclassicality coherence}) has the form of Eq. (\ref{product bases of two qubit}) with $\alpha_1=\alpha_2=\pi/2$ and $\beta_1=\beta_2=3\pi/4$, so that $\{\ket{b_*}\}=\{\ket{b_{*1+},b_{*2+}},\ket{b_{*1+},b_{*2-}},\ket{b_{*1-},b_{*2+}},\ket{b_{*1-},b_{*2-}}\}$, where $\ket{b_{*k\pm}}=\frac{1}{\sqrt{2}}(\ket{0}_k\pm e^{i3\pi/4}\ket{1}_k)$, $k=1,2$, hence, it is mutually unbiased with the incoherent basis $\{\ket{a}\}$. One can further check that the state is also maximally coherent with respect to $\{\ket{b_*}\}$, namely, it can be expressed as $\ket{\psi}=(-e^{-i\pi/4}\ket{b_{*1+},b_{*2+}}+\ket{b_{*1+},b_{*1-}}+\ket{b_{*1-},b_{*2+}}+e^{-i\pi/4}\ket{b_{*1-},b_{*2-}})/2$, in accord with Proposition 3. For completeness, the associated KD quasiprobability written in the form of $4\times 4$ matrix reads
\begin{eqnarray}
&&\{{\rm Pr}_{\rm KD}(\varrho)\}_{ab_*}\nonumber\\
&=&
\begin{pmatrix} 
-\frac{221}{2500}(1+i) & \frac{1}{8}& \frac{1}{8}& \frac{221}{2500}(1+i)\\
\frac{1}{8} & \frac{221}{2500}(1-i)& -\frac{221}{2500}(1-i)& \frac{1}{8}\\
\frac{1}{8} & -\frac{221}{2500}(1-i)& \frac{221}{2500}(1-i)& \frac{1}{8}\\
\frac{221}{2500}(1+i) & \frac{1}{8}& \frac{1}{8}& -\frac{221}{2500}(1+i)
\end{pmatrix}.
\end{eqnarray} 
One can see in the above example that the coherence is captured by both the nonreal values and the negative values of the real part of the KD quasiprobability. 

Next, let us discuss another example of two-qubit coherent state whose associated KD quasiprobability is real but negative. Assume that the state is maximally coherent relative to the orthonormal product basis $\{\ket{a}\}=\{\ket{00},\ket{01},\ket{10},\ket{11}\}$ as $\ket{\psi}=(\ket{00}+\ket{01}+\ket{10}-\ket{11})/2$. Then we obtain numerically $C_{\rm KD}^{\rm NCl}(\ket{\psi}\bra{\psi};\{\ket{00}\bra{00},\ket{01}\bra{01},\ket{10}\bra{10},\ket{11}\bra{11}\})=\sqrt{4}-1=1$ in accord with Proposition 3. This can again be checked analytically using Eq. (\ref{KD-nonclassicality coherence is equal to the measurement uncertainty for pure states}). Moreover, the second orthonormal product basis which achieves the supremum in Eq. (\ref{KD-nonclassicality coherence}) is given by ${\ket{b_*}}=\{\ket{+,+},\ket{+,-},\ket{-,+},\ket{-,-}\}$, $\ket{\pm}=\frac{1}{\sqrt{2}}(\ket{0}\pm\ket{1})$, i.e., when the parameters in Eq. (\ref{product bases of two qubit}) are given by $\alpha_1=\alpha_2=\pi/2$ and $\beta_1=\beta_2=0$. One can check that $\{\ket{b_*}\}$ is mutually unbiased with the incoherent basis $\{\ket{a}\}$, and moreover, the state is also maximally coherent relative to $\{\ket{b_*}\}$, in accord with Proposition 3. Finally, the associated KD quasiprobability written in the form of $4\times 4$ matrix reads
\begin{eqnarray}
&&\{{\rm Pr}_{\rm KD}(\varrho)\}_{ab_*}=
\begin{pmatrix} 
\frac{1}{8} & \frac{1}{8}& \frac{1}{8}& -\frac{1}{8}\\
\frac{1}{8} & -\frac{1}{8}& \frac{1}{8}& \frac{1}{8}\\
\frac{1}{8} & \frac{1}{8}& -\frac{1}{8}& \frac{1}{8}\\
-\frac{1}{8} & \frac{1}{8}& \frac{1}{8}& \frac{1}{8}
\end{pmatrix},
\label{KD quasiprobability of maximally coherent state with real but negative values}
\end{eqnarray}
with all real elements. Hence, in this case, the coherence in the maximally coherent state is captured entirely by the total negativity of the associated KD quasiprobability.

\end{document}